\documentclass[twocolumn]{aastex701}
\usepackage{adjustbox}

\begin{document}

\title{Stars Born in the Wind II: Widespread Extra-planar Star Formation in M82's Halo}

\author[0000-0002-8406-0136]{Vaishnav V. Rao}
\affiliation{Department of Astronomy, University of Michigan, 1085 S. University Ave, Ann Arbor, MI 48109-1107, USA}
\email[show]{vvrao@umich.edu}

\author[0000-0002-5564-9873]{Eric F.\ Bell}
\affiliation{Department of Astronomy, University of Michigan, 1085 S. University Ave, Ann Arbor, MI 48109-1107, USA}
\email{ericbell@umich.edu}

\author[0000-0003-2599-7524]{Adam Smercina}\thanks{NHFP Hubble Fellow}
\affiliation{Space Telescope Science Institute, 3700 San Martin Dr., Baltimore, MD 21218, USA}
\email{asmercina@stsci.edu}

\author[0009-0001-2638-8723]{Elliott Besirli}
\affiliation{Department of Astronomy, University of Michigan, 1085 S. University Ave, Ann Arbor, MI 48109-1107, USA}
\email{ebesirli@umich.edu}

\author[0000-0002-8084-8612]{Andrew Dolphin}
\affiliation{Raytheon, Tucson, AZ 85756, USA}
\affiliation{Steward Observatory, University of Arizona, Tucson, AZ 85721, USA}
\email{adolphin@rtx.com}

\author[0000-0003-2325-9616]{Antonela Monachesi}
\affiliation{Departamento de Astronom\'{i}a, Universidad de La Serena, Avda. R\'{a}ul Bitr\'{a}n 1305, La Serena, Chile}
\email{amonachesi@userena.cl}

\author[0000-0002-7502-0597]{Benjamin Williams}
\affiliation{Department of Astronomy, University of Washington, Box 351580, Seattle, WA 98195-1580, USA}
\email{benw1@uw.edu}

\author[0000-0002-1264-2006]{Julianne J.\ Dalcanton}
\affiliation{Center for Computational Astrophysics, Flatiron Institute, 162 Fifth Avenue, New York, NY 10010, USA}
\affiliation{Department of Astronomy, University of Washington, Box 351580, Seattle, WA 98195-1580, USA}
\email{jdalcanton@flatironinstitute.org}

\author[0000-0001-6982-4081]{Roelof S. de Jong}
\affiliation{Leibniz-Institut f\"{u}r Astrophysik Potsdam (AIP), An der Sternwarte 16, 14482 Potsdam, Germany}
\email{rdejong@aip.de}

\begin{abstract}
Galaxies evolve in tandem with their environments--- mergers and gas inflows drive galaxy growth while galactic outflows launched by supernovae may seed the galactic environment with gas, metals, and energy, fueling star-formation far from the main bodies of galaxies. The formation histories of young stars in the stellar halos of nearby galaxies can help understand this interplay. We thus present the most detailed map to date of young stars in the stellar halo of M82, a starburst galaxy in the M81 Group that hosts a prototypical outflow, using Hubble Space Telescope (HST) and Subaru Hyper-Suprime Cam observations. We find widespread extraplanar populations of stars with ages $\lesssim630\,$Myr, with clear detections of stars up to $\sim5\,$kpc to the south in unique arc-like stellar features (Southern Arcs) and in a new stellar trail up to $\sim20\,$kpc to the east (M82's Tail), originating from the Southern Arcs. We estimate a total halo star formation of $\sim4\times10^6\,M_\odot$ in the last $630\,$Myr. Overall, the star formation history (SFH) of the M82 Tail is correlated with periods of heightened star cluster formation in the M82 disk, which suggests the influence of the starburst outflow. Further, the fraction of young stars decreases as we move away from M82 to the east. We forward a picture where the M82 Tail formed from ram pressure stripped gas arising from M82's westward motion, triggered by shocks from the outflow. 

\end{abstract}

\keywords{\uat{Starburst galaxies}{1570} --- \uat{Circumgalactic medium}{1879} --- \uat{Galaxy winds}{626} --- \uat{Star formation}{1569} --- \uat{Stellar Halos}{598}}


\section{Introduction}
The evolution of galaxies is deeply influenced by their surrounding environment, with interactions and mergers playing a pivotal role. In the framework of the $\Lambda$-Cold Dark Matter ($\Lambda$CDM) paradigm, galaxies grow hierarchically, driven by inflows of gas from cosmic filaments and frequent mergers with other galaxies \citep[e.g.,][]{W&R1978, Bullock2001}. These mergers profoundly reshape galaxies, altering their structure, star formation rates, and overall morphology \citep[e.g.,][]{Toomre&Toomre1972,Cox2008}. The interplay between a galaxy and its environment, however, is not a one way street. Accreting supermassive black holes (BHs; or active galactic nuclei) and/or supernovae explosions following intense star formation episodes within a galaxy inject energy and metals into the surrounding interstellar medium (ISM), powering galaxy-scale outflows \citep{C&C1985}. These outflows exert significant ``feedback" on the ISM and the circumgalactic medium (CGM), driving shocks, transporting gas, and influencing subsequent generations of star formation \citep[e.g.,][]{Veilleux2005, Borthakur2013}.

The CGM and stellar halos of nearby galaxies have proved to be valuable tools for studying the interplay between galaxies and their environments. On the one hand, multi-wavelength observations of the CGM tracing for example neutral hydrogen (HI) in radio, molecular gas in the submillimeter, dust in the infrared, and hot ionized gas in X-rays reveal contemporary insights into the multi-phase gas surrounding a galaxy. While on the other hand, the stellar halo of the galaxy--- tenuous swaths of stars strewn across its outskirts--- typically serve as fossil records for ancient merger events, preserving clues in their ancient red giant branch (RGB) or asymptotic giant branch (AGB) stellar populations \citep[e.g.,][]{Ibata2014,Williams2015,Okamoto2015, Smercina2020, Velguth2024}. While individual stars in the main bodies of nearby galaxies are challenging to study in detail owing to severe crowding, dust obscuration, and rapid dynamical mixing, those in the stellar halo are largely free from these complications. Only a handful of nearby stellar halos analyzed, however, preserve a more recent archaeological record by hosting young and intermediate age ($\lesssim1\,$Gyr) stellar populations \citep[e.g., LMC, Milky Way, Centaurus A, M81][]{Moni-Bidin2017, Price-Whelan2019, Rejkuba2002, Makarova2002}. These younger populations open a window into an intermediate epoch of galaxy evolution otherwise inaccessible through CGM or older stellar population studies. 

There are multiple ways through which young stars can come to populate a stellar halo of a galaxy. Stars can form directly within the dense, clumpy molecular gas phases of galactic outflows \citep[e.g,][]{Maiolino2017, Gallagher2019, Yu2020}, they can form within CGM clouds shocked by outflows \citep[e.g., Centaurus A and Teacup Galaxy;][]{Rejkuba2002, GK&W2010, Crockett2012, Venturi2023}, they can form within gas stripped from the main body of the galaxy due to tidal forces \citep[e.g., Milky Way, M81;][]{Price-Whelan2019,Makarova2002, S&M2001, Okamoto2019} or ram pressure forces \citep[e.g.,Virgo cluster galaxy IC3418 and other jellyfish galaxies;][]{Kenney2014, Ebeling2014}, or they may be scattered into the stellar halo from the disk as a result of dynamical processes \citep[e.g.,][]{Dorman2013, Johnston2017, Seth2005, Font2011, Cooper2015, Zolotov2009}. Studying the formation histories of these young stars in the context of a galaxy's environment and evolutionary history can help us fill in the pieces of temporal evidence needed for understanding how gravity and feedback influence gas flows across galactic scales and cosmic times.

M82, within its immediate cosmological neighbourhood--- the M81 Group, serves as an unprecedented laboratory for studying the interplay between a galaxy and its environment. Situated at a distance of 3.6\,Mpc \citep{Dalcanton2009}, this edge-on galaxy hosts a prototypical nuclear starburst, which drives strong, multiphase, bipolar outflows (``superwinds") along its minor axis \citep{L&S1963, O&M1978, Bland&Tully1988}. The starburst activity is widely attributed to a close encounter with Milky Way-mass M81 within the last $\sim1$\,Gyr \citep[e.g.,][]{deGrijs2001, RM2011, Mayya2006, Li2015}. Ongoing interactions with both M81 and the LMC-like NGC 3077 further complicate its environment, generating significant tidal debris composed of streams and filaments of HI gas \citep{Yun1994, Okamoto2015, deBlok2018, Smercina2020}. Tidal disruption of the outer M82 disk has resulted in large filaments of HI such as the North-East Spur and the Southwest Spur, which are anchored to the edges of the disk \citep{Yun1993}. In short, M82's CGM is chaotic, littered with material hurled by M82's starburst outflow and gas stripped from M82's disk.

Being one of the closest ensembles of galaxies with evidence of ongoing galaxy-galaxy interactions, the stellar halo of the M81 Group has been mapped in exquisite detail by the Subaru Hyper-Suprime Cam \citep[e.g.,][]{Okamoto2015, Smercina2017}. Around M82 in particular, decades of studies have mined the stellar content of its halo using facilities such as the Canada-France-Hawaii Telescope and the Hubble Space Telescope (HST). These studies have collectively revealed a widespread distribution of young stars around M82 along with striking arc-like stellar features known as the Southern Arcs, located $\sim5\,$kpc South of the M82 disk \citep[e.g.,][]{Davidge2008b, Sun2005, Davidge2008a}. More recently, using data from the ACS Nearby Galaxy Survey Treasury \citep[ANGST;][]{Dalcanton2009} and new deep HST Wide Field Camera 3 (WFC3) observations of the Southern Arcs, \citet{Rao2025} mapped the $\lesssim400\,$Myr stars around M82 and demonstrated using resolved-star star formation history (SFH) measurements that the Southern Arcs formed directly due to the influence of M82's starburst outflow $\sim100\,$Myr ago. 

There is, however, more to be learned from M82's unique stellar halo. The widespread distribution of extraplanar young stars in M82 stands out in the sense that halo star formation in other galaxies is often localized in clumps of stripped or shocked gas \citep[e.g., Cen A, LMC, Teacup;][]{Crockett2012, Price-Whelan2019, Venturi2023}. Could substantial circumgalactic star formation be a feature of powerful starburst systems? M82’s proximity allows us to test this idea directly by mapping young stellar features in unprecedented detail and probing their origins. Additionally, ubiquitous young halo populations in nearby systems such as M82 may indicate the importance of starburst outflows as a mechanism for triggering \textit{in-situ} halo star formation--- especially in the context of the high-redshift Universe, where starbursts were more common. To that end, the aim of this paper is to build on the work of \citet{Rao2025}, map the striking widespread distribution of young stars in M82's halo by augmenting a panoramic ground-based resolved star dataset, and comprehensively study halo star formation histories with the goal of understanding the co-evolution of M82 and its broader environment. 

The structure of the paper is organized as follows. In Section \ref{sec:subaru}, we highlight a new trail of young stars to the South-East of M82, originating from the Southern Arcs and henceforth referred to as the ``M82 Tail", using panoramic resolved star maps from Subaru HSC. In Section \ref{sec:hst-deep}, we present deep follow-up HST observations of stars along the M82 Tail and derive their formation histories. In Section \ref{sec:ANGST}, we validate the young halo stars in the ANGST footprint against stars detected using the James Webb Space Telescope's (JWST) Near Infrared Camera (NIRCam). In Section \ref{sec:discussion}, we place these observational results in the context of M82's evolutionary history and discuss the interplay between the starburst outflow and stripped gas in its surroundings. Finally, we propose observational tests to validate our findings and summarize our results in Section \ref{sec:summary}. 

\begin{figure*}
    \centering
    \includegraphics[width= \linewidth]{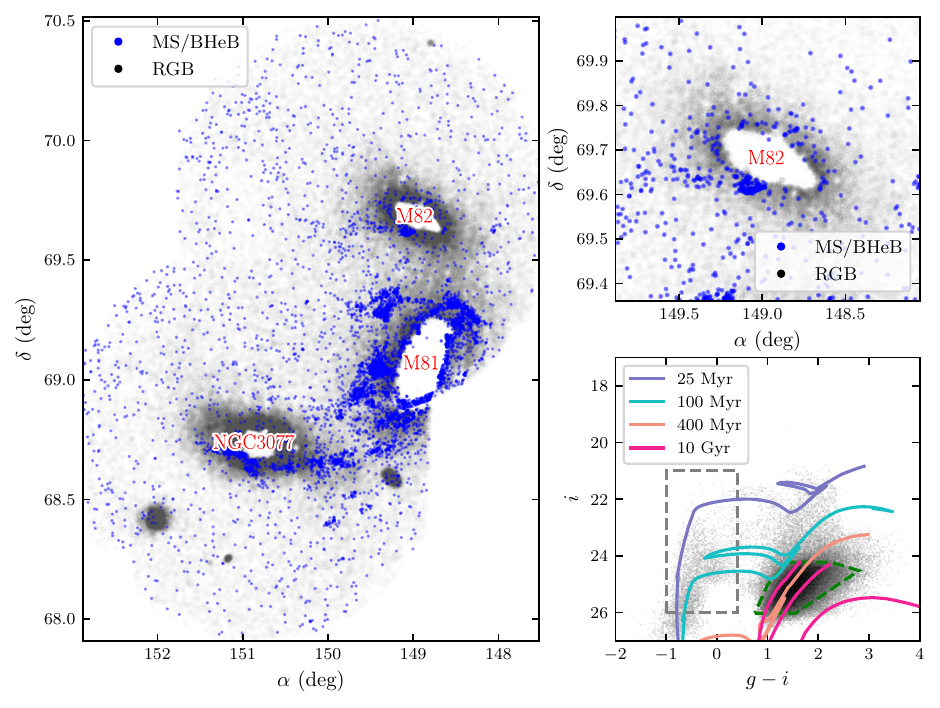}
    \caption{(\textbf{Left}) A Subaru HSC panoramic view of the M81 Group in resolved stars. Black points represent RGB stars and blue points represent young MS and BHeB stars. (\textbf{Top Right}) Zooming into a 40\,kpc box around M82, the Southern Arcs and the M82 Tail consisting of $<400\,$Myr stars targeted in this work can be seen extending eastwards from M82. (\textbf{Bottom Right}) CMD of sources classified as stars in the Subaru HSC photometry. PARSEC \citep{Bressan2012-PARSEC} isochrones (shifted assuming $(m-M)_0 = 27.8$) are over-plotted to broadly indicate the regions occupied by the $<400\,$Myr MS and BHeB stars (gray dashed line) and the ancient, RGB stars (dashed green line). These RGB stars are represented by $10\,$Gyr isochrones ([M/H] = 0.0, -1.0, -2.0).}
    \label{fig:m81-subaru}
\end{figure*}

\section{Subaru Panoramic View of M82's Halo}\label{sec:subaru}
In this Section we present a panoramic view of young, resolved stars in the M81 Group using Subaru HSC, similar to that presented in \citet{Okamoto2019}. We use the same dataset as \citet{Smercina2017}, but implement an improved method for star-galaxy separation, which reveals a diffuse trail of young stars to the southeast of M82--- the M82 Tail.

\subsection{Subaru: Observations}
The M81 Group was observed with Subaru HSC through the Gemini-Subaru exchange program (PI: Bell, 2015A-0281). The Group was imaged in the ``classical" observing mode between the nights of March 26 and 27, 2015. The survey consisted of two 1.5$^{\circ}$ field-of-view (FOV) pointings in each of the three ($g$, $r$, $i$) filters. These pointings were primarily chosen to cover the outer regions of all three major interacting galaxies in the Group--- M81, M82, and NGC 3077. Further observing details such as integration times for each field and filter combination can be found in Table 1 of \citet{Smercina2020}. 

We reduced the imaging data following \citet{Smercina2017}, using the HSC optical imaging pipeline \citep{Bosch2018}. The pipeline performs photometric and astrometric calibration using the Pan-STARRS1 catalog \citep{Magnier2013} and reports the final magnitudes in the HSC natural system, which we correct to the SDSS filter system. The version of the pipeline adopted for this dataset performs background subtraction using an aggressive 32 pixel mesh, optimizing point-source detection and eliminating most diffuse light. We detect sources in all three bands, but prioritize the $i$-band to determine reference positions for forced photometry. We then perform forced photometry on sources in the $gri$ coadded image stack.

We corrected all magnitudes for galactic foreground extinction using the Planck Collaboration's GNILC dust map \citet{Planck2016} through the \texttt{dustmaps} software package \citep{dustmaps}. The $E(B-V)$ throughout the M81 Group appears relatively constant at $\simeq0.1$. However, the innermost regions of M82 suffer from artificially high estimated extinction due to ``contamination" from dust emission, which is why we limit $E(B-V)$ to a maximum of 0.1. We found that the image depth was uniform across the two fields, yielding extinction corrected point source detection limits of $g=27$, $r=26.5$, and $i=26.1$, measured at $\sim5 \sigma$. Seeing was relatively stable, resulting in consistent point source function (PSF) sizes of 0.7\arcsec-0.8\arcsec down to the detection limits.

\subsection{Subaru: Star-Galaxy Separation}
Stars resolved by Subaru HSC at the distance of the M81 Group have magnitudes of $i>24$. At these magnitudes, resolved faint stars are easily confused with faint, background galaxies, fundamentally limiting the accuracy of wide-area ground based survey datasets. At brighter magnitudes, one can separate stars from background galaxies by morphology alone \citep[choosing galaxies to be unresolved; e.g.,][]{Bertin_Arnouts96, Desai12, Slater20}. However, morphological cuts alone misclassify as stars many unresolved faint background sources \citep[e.g.,][]{Okamoto2015, Crnojevic2016}. If the dataset includes three or more passbands, color-color information can be used to reduce background galaxy contamination, either by using explicit color-color cuts \citep[e.g.,][]{Baldry2010, Smercina2017, Smercina2020} or by using supervised machine learning techniques \citep[e.g.,][]{Soumagnac2015, Bell2022, Bell2025}. 

The star-galaxy separation for the Subaru M81 dataset presented in this work follows that of \citet{Bell2025}; please refer to that source for more details. The M81 Group pointings contain a number of HST fields that offer a sizable sample of stars with matches in Subaru to train supervised machine learning classifiers. We choose to focus on magnitudes, colors and how extended sources are compared to the local PSF size in multiple bands as features that offer a combination of discriminating power and the ability to generalize outside of the tiny areas covered by HST. The pipeline output that we use reports covariance matrices of the 2nd moment of a source's light in the $xx$, $yy$, and $xy$ directions (where $x$ is aligned with RA and $y$ is aligned with Dec). Here, we choose to focus on the difference in sizes between a source and estimate of local PSF size from the HSC pipeline, giving six morphological features: $\sigma_{xx,g}-\sigma_{xx,g,PSF}$, $\sigma_{yy,g}-\sigma_{yy,g,PSF}$, $\sigma_{xx,r}-\sigma_{xx,r,PSF}$, $\sigma_{yy,r}-\sigma_{yy,r,PSF}$, $\sigma_{xx,i}-\sigma_{xx,i,PSF}$ and $\sigma_{yy,i}-\sigma_{yy,i,PSF}$. These differences in sizes are more robust to variations in data quality between fields, and within a field, than the sizes themselves. To these we add three more features: $i$-band PSF-fit magnitude, and PSF-fit derived $g-r$ and $r-i$ colors. We rescale each feature to have a `robust' standard deviation (a scaled median absolute deviation; using \texttt{astropy}'s \texttt{mad\_std}) $\sigma_{robust} = 1.0$. We use $K$-nearest neighbors classification, with a majority classification from the 15 nearest neighbor in this 9-dimensional space, for classification. Similar results would result from other classification methods (see also \citealt{Bell2025}). 
We use half of the sources for training, and use the rest as a test set to quantify classifier performance.


Our Subaru dataset contains 6458 stars identified with HST. To build a balanced classifier, we drew 29045 “background” sources directly from Subaru imaging, sampled from a large region well away from M81. This set includes both foreground stars --- some of which are not stars of interest --- and background galaxies. The number of background sources was chosen to match the size and composition of the raw Subaru dataset, which contains a somewhat higher fraction of background sources. As expected, the ratio of stars to background sources affects classifier performance: increasing the background fraction reduces stellar completeness, since stronger evidence is then required to classify an object as a star. However, we find that even large changes in this balance do not significantly alter our conclusions. We chose a background-heavy balance that somewhat favors low contamination over completeness, resulting in a slightly smaller but purer sample of stars.

To preserve faint halo features while limiting contamination, we selected a star–background balance and nearest-neighbor parameter that reduce false positives without sacrificing too much stellar completeness. In the test set, which includes half of the HST-identified stars (excluding bright foreground stars), the classifier recovers 2122/3229 stars (66\% recall), while misclassifying the other 1107 stars as background. At the same time, only 498 background objects (out of the total of 14522 background objects) are incorrectly identified as stars, yielding a stellar precision of 81\% (real stars are 2122 of the 2620 estimated `stars' in the test set). Overall, the classifier correctly rejects 14024/14522 background objects (97\%), substantially reducing contamination in the final star sample.

In what follows, when we are discussing Subaru data, we will refer to those objects classified as stars by this $K$-nearest neighbors technique as `Subaru stars'. 

\subsection{Subaru: Young and Intermediate Age Star Map of M82's halo}
A panoramic map of young $\lesssim 400\,$Myr Subaru stars in the M81 Group reveals a diffuse trail extending due East from the South of M82 (Figure \ref{fig:m81-subaru}). We select these blue stars from the extinction-corrected $g-i$ vs. $i$ color-magnitude diagram (bottom right panel, Figure \ref{fig:m81-subaru}). For context, we also include in the map the ancient $\sim10\,$Gyr red giant branch (RGB) stars that dominate the stellar halo at these depths \citep{Smercina2017, Smercina2020, Okamoto2015}. Significant over-densities of RGB stars are detected around M81, M82, and NGC 3077, including around smaller dwarf satellites such as KDG61, IKN, and BK5N. Our map of young stars recovers many of the features also detected by \citet{Okamoto2019} such as the Arp's Loop, Holmberg IX, the Garland, and the Southern Arcs of M82. The feature we are interested in--- the M82 Tail--- appears immediately to the East of the Southern Arcs and is not noticeably associated with a corresponding over-density in the RGB stars. 

\begin{deluxetable}{lccccccc}
\tablecaption{HST Observations of M82's Halo\label{tab:obs}}
\tablehead{
\colhead{Field} & \colhead{RA} & \colhead{Dec} & \colhead{Instrument} & \colhead{Filter} & \colhead{Exposure Time (s)} & \colhead{50\% Completeness}
& \colhead{GSTs}}
\startdata
WFC3-2 & 09:57:30.947 & +69:37:40.97 & WFC3 & F475W & 3482 & 28.1 & 1125\\
 &  &  &  & F814W & 6964 & 26.5 & \\
WFC3-3 & 09:58:46.348 & +69:38:13.54 & WFC3 & F475W & 4351 & 28.1 & 468\\
 &  &  &  & F814W & 8702 & 26.5 & \\
ACS-1 & 09:57:13.403 & +69:40:22.85 & ACS & F475W & 3363 & 28.1 & 7633\\
 &  &  &  & F814W & 6816 & 26.5 & \\
ACS-2 & 09:58:23.355 & +69:41:12.92 & ACS & F475W & 3363 & 28.1 & 1729\\
 &  &  &  & F814W & 6816 & 26.5 & \\
ACS-3 & 09:59:36.580 & +69:41:59.52 & ACS & F475W & 3363 & 28.1 & 872\\
 &  &  &  & F814W & 6816 & 26.1 & \\
\enddata
\tablecomments{All observations were taken between January 2021 and February 2022}
\end{deluxetable}

\section{Deep HST Halo Fields}\label{sec:hst-deep}
\begin{figure*}
    \centering
    \adjustbox{max height=\dimexpr\textheight-10\baselineskip\relax}{%
        \includegraphics{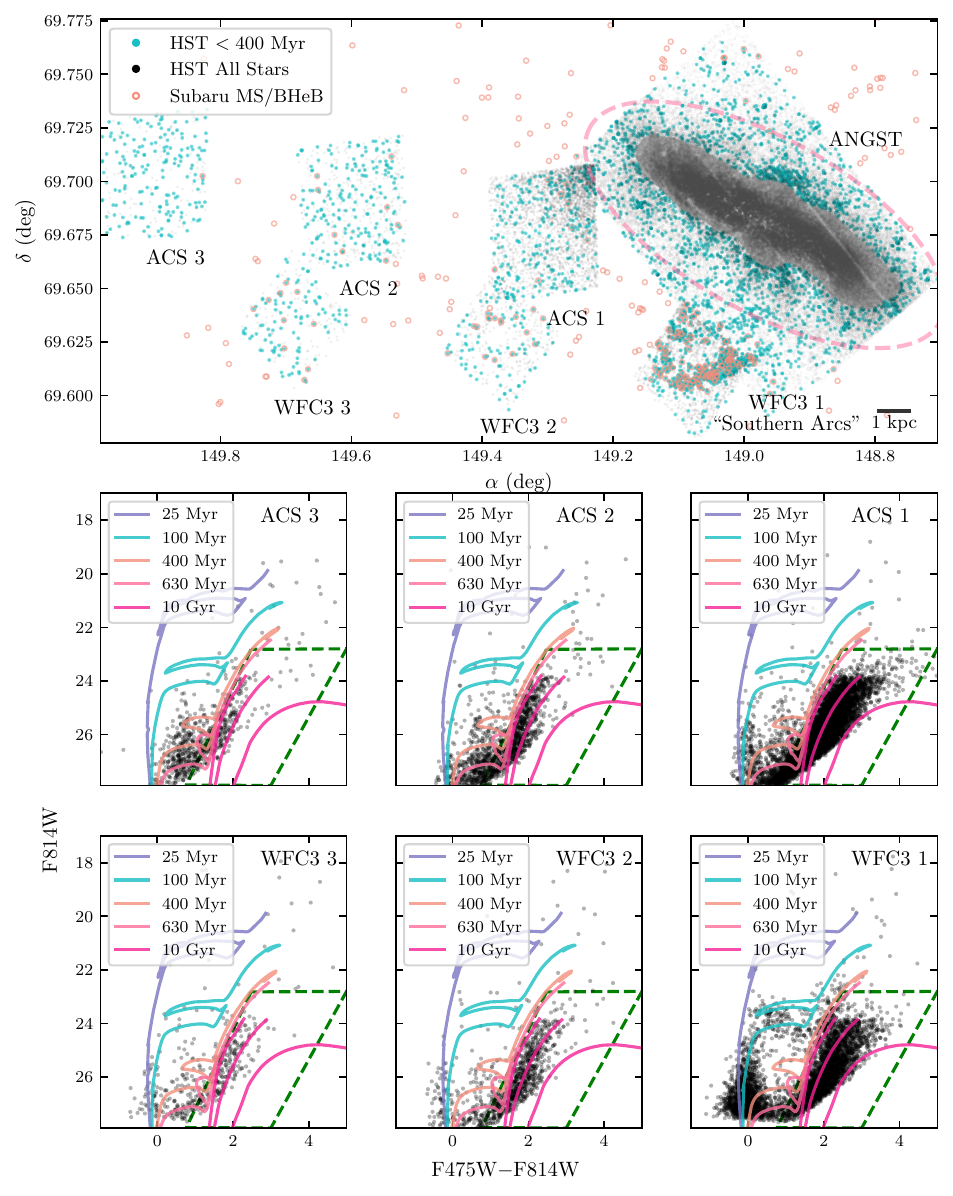}
    }
    \caption{(\textbf{Top}) An HST resolved star view of M82 and it's halo with stars from the ANGST dataset \citep{Dalcanton2009}, the Southern Arcs field \citep{Rao2025}, and this work. The points in cyan represent populations of stars $\lesssim400\,$Myr, selected by excluding the RGB stars from the CMD of each dataset (dashed green regions in the CMDs below) with depths matched (F814W $<26$). The points in salmon represent the MS and BHeB stars from Subaru HSC. The M82 disk has been masked out to accentuate structures that follow the breakout of the outflow from the disk. The pink dashed line represents the approximate crowding limit for the Subaru dataset. (\textbf{Bottom}) CMDs of `GST' sources from the Deep Halo Fields (including the Southern Arcs) arranged to reflect their relative positions on the sky. PARSEC \citep{Bressan2012-PARSEC} isochrones with foreground extinction (shifted assuming $A_V= 0.2$ and $(m-M)_0 = 27.8$) are over-plotted to broadly explain the multiple stellar populations in each field. The $10\,$Gyr isochrones ([M/H] = 0.0, -1.0, -2.0) correspond to the old, RGB halo stars. For the $\leq630\,$Myr isochrones, we assume an [M/H]= -1.0.}
    \label{fig:m82-mosaic}
\end{figure*}

In this Section, we present deep, follow-up observations of stars in the M82 Tail using HST (Figure \ref{fig:m82-mosaic}) and derive their formation histories. The HST fields pointed along the M82 Tail will henceforth be referred to as the ``Deep Halo Fields".

\subsection{Deep Halo Fields: Observations}
The Deep  Halo Fields presented in this paper were observed through the Hubble Space Telescope (HST) Cycle 28 program GO-16185 (PI: Adam Smercina) between January 2021 and February 2022.  We targeted parts of M82's halo directly south of the disk and towards the east with HST’s UVIS channel of the Wide Field Camera 3 (WFC3), and with coordinated parallel observations using HST's Wide Field Channel (WFC) of the Advanced Camera for Surveys (ACS), for a total of 12 orbits. The Southern Arcs field described in \citet{Rao2025} is part of these observations.

In order to maximize scheduling, the 12 orbits were organized into 12 individual visits, with four visits assigned to each WFC3+ACS pointing. Exposures in F475W and F814W were distributed across each set of four visits to optimize the detection of sources with main-sequence stellar colors, resulting in an approximate 33/67 split in exposure time. A 3\degr ORIENT range (182–185\degr) was specified for the first visit of each target to allow maximum schedulability while avoiding overlaps between ACS parallel fields and subsequent WFC3 prime pointings. To achieve equivalent depth in both the prime (WFC3) and parallel (ACS) observations, each subsequent visit for a given target was executed at the same ORIENT as the first visit.

For each Deep Halo Field, three of the four visits (Visit 2–4), which contained exposures in only a single filter, employed three-point dither patterns with 2.98\arcsec spacing to provide $\sim1/2$ pixel sub-sampling of the PSF. One visit per field (Visit 1) combined F475W and F814W exposures with WFC3. These visits included two F814W dithers separated linearly by 2.98\arcsec and offset from the target center by 0.98\arcsec in both X and Y, along with a single F475W exposure. The combined dither pattern for both filters fully sampled the WFC3 and ACS chip gaps, enabled robust cosmic-ray rejection, and, together with the sub-pixel dither spacing, produced a well-sampled PSF. The field of view and plate scale for WFC3/UVIS is 162\arcsec$\times$162\arcsec and 0.04\arcsec/pixel, while that for ACS/WFC is 202\arcsec$\times$202\arcsec and 0.05\arcsec/pixel. The total exposure time and 50\% completeness depth (measured from artificial star tests; Section \ref{sec:ast}) in each filter are summarized in Table~\ref{tab:obs}. These HST data can be found in MAST: \dataset[10.17909/7vae-tz90]{http://dx.doi.org/10.17909/7vae-tz90} 

\subsection{Deep Halo Fields: Photometry and ASTs}\label{sec:ast}
We performed point spread function (PSF) fitting on the HST pipeline-calibrated images (\texttt{flc} extension) using the latest version of the \texttt{DOLPHOT} software package \cite{Dolphin2000PHOT, Dolphin2016}. We conducted artificial star tests (ASTs) to evaluate the photometric quality, completeness, and bias of our observations. We generated input lists of 250,000 artificial stars per field, sampled from a realistic range of stellar population models employing the Bayesian Extinction and Stellar Tool (\texttt{BEAST}) package \citep{Gordon2016}. We then injected these artificial stars into the images and recovered them through \texttt{DOLPHOT} following the methodology established by the Panchromatic Hubble Andromeda Treasury (PHAT) program \citep{Williams2014}.

Using the results of the ASTs, we computed completeness curves for each filter. These curves help us determine the optimal selection criteria for stellar sources. We classified sources as `good stars' (GST) if they met our criteria for \texttt{DOLPHOT}'s signal-to-noise ratio (\texttt{SNR}), crowding (\texttt{CROWD}), and sharpness (\texttt{SHARP}) parameter outputs. In the uncrowded halo fields, the \texttt{CROWD} parameter was particularly useful for identifying spurious sources detected along the diffraction spikes of bright foreground stars and other photometric artifacts. We chose the smallest \texttt{CROWD} parameter that effectively removes most of these contaminating sources without significantly impacting photometric completeness. The final criteria for selecting GSTs were: \texttt{SNR} $> 4$, \texttt{SHARP}$^2 < 0.2$, and \texttt{CROWD} $< 0.25$. We list the number of GSTs obtained in each field after implementing the quality cuts in Table \ref{tab:obs}. Please refer to Appendix \ref{appendix:comp-bias} for completeness curves and photometric biases.

\subsection{Deep Halo Fields: Star Formation Histories}
Much of our star formation history (SFH) analysis closely follows the procedure of \citet{Rao2025}. We derived the SFHs of the Deep Halo Fields using the color-magnitude diagram (CMD) fitting code \texttt{MATCH} \citep{Dolphin2002, Dolphin2012, Dolphin2013}. \texttt{MATCH} determines the linear combination of single-burst stellar populations that best reproduces the observed CMD after incorporating photometric biases, completeness from ASTs, and dust extinction. The best-fit SFH is then established using a Poisson maximum likelihood technique \citep[see][for more details]{Dolphin2002, Dolphin2012, Dolphin2013}, which compares the star counts in each CMD bin between the data and the model.

The CMDs of the Deep Halo Fields lack the depth required to resolve the oldest main-sequence turnoff and therefore cannot accurately constrain the ages of the oldest stars \citep[e.g.,][]{Williams2017, Weisz2014}. Since our focus is on the younger populations, we excluded the CMD regions occupied by the old RGB stars \citep[see, e.g.,][]{Lewis2015,Lazzarini2022} using the \texttt{exclude\_gates} functionality of \texttt{MATCH} and restricted the fits to regions sensitive to populations formed within the last $\sim630$\,Myr. These age-sensitive regions are home to blue and red core helium burning stars (BHeB and RHeB, respectively), which serve as robust chronometers for recent star formation episodes, as their distinct magnitudes and colors allow precise age determinations \citep{Dohm-Palmer2001, Weisz2008, McQuinn2012}. 

The input parameters for our SFH fits were: distance modulus, IMF, extinction, metallicity, binary fraction, magnitude limits, and age bins. We assumed that the younger populations in all halo fields reside at approximately the same distance as the ancient population in the Southern Arcs field and adopted a distance modulus of 27.8 \citep[TRGB distance for this field by][we also test for the effect of $\pm$0.2\,mag variations around that adopted distance modulus]{S&S2024}, which is also the distance modulus of M82. We set the magnitude limits to 50\% completeness from ASTs, fixed the binary fraction at 0.35, and adopted a \citet{Kroupa2001} IMF. We used logarithmic age bins from 6.6 to 8.8 (approximately 4–630\,Myr) with a spacing of 0.2 dex. These are particularly coarse time bins as the deep halo fields have on average $\sim600$ GST sources after excluding the RGB and coarse time bins reduce the random uncertainty in star formation rates (SFRs) arising from detecting fewer stars in CMD regions. Combined with the RGB exclusion, this binning strategy isolates the potentially distinct evolutionary pathways of the young MS and core HeB stars, separate from the underlying halo RGB population. To avoid unphysical chemical histories, we enforced a monotonically increasing age-metallicity relationship over our age range using the \texttt{zinc} option to capture enrichment from star formation --- a recommended strategy \citep[e.g.,][]{Weisz2008}.

To assess the effect of stellar evolution models on the inferred SFHs, we carried out \texttt{MATCH} fits with PADUA (or Padova) isochrones \citep{Marigo2008} including updated AGB tracks \citep{Girardi2010}, MIST isochrones \citep{Dotter2016-MIST}, and the updated BASTI set \citep{Hidalgo2018-BASTI}. The PADUA models span [M/H] = [$-$2.3, 0.1], the MIST models [$-$2.0, 0.5], and the BASTI models [$-$3.2, 0.4], all with 0.1 dex resolution. We restricted the metallicity range of the oldest ($\sim630$\,Myr) time bin to [$-$2.0, $-$0.5] and that of the youngest ($\sim4$\,Myr) time bin to [$-$1.0, $-$0.2] and required the mean metallicity to increase with time.

We also included extinction due to both foreground dust and the dust local to M82. Foreground extinction is parametrized by $A_V$, which reddens all stars uniformly, while differential extinction $dA_V$ accounts for the patchy dust distribution within M82. This simple model is well-suited for young populations (age $<1$\,Gyr), which typically experience a top-hat–like distribution of reddening \citep[e.g.,][]{Dolphin2003, Weisz2014, Lazzarini2022} In \texttt{MATCH}, extinction is applied to all stars in a uniform distribution between $A_V$ and $A_V + dA_V$. Thus, $A_V$ shifts the entire CMD fainter and red-ward, whereas $dA_V$ broadens and dims the main sequence (MS). We explicitly set $dA_{Vy}=0$ to ensure no additional differential extinction is added to stars $<100\,$Myr. 

We determined the best values for $A_V$ and $dA_V$ both by visually inspecting the model Hess diagram (2D histogram of the CMD) features and by minimizing the fit parameter obtained by running MATCH multiple times on a grid of $A_V$ and $dA_V$. The grid allowed $A_V$ to range between 0.05 and 1.0 in steps of 0.05 and $dA_V$ to range between 0.2 and 0.8 in steps of 0.2. For WFC3 Fields 2 and 3 and ACS Field1, which have a prominent MS, $dA_V$ of 0.4 produced the best fit model Hess diagram. This particular $dA_V$ also produced model Hess diagrams that visually had similar MS widths to the observed Hess diagram. However, for ACS Fields 2 and 3, which do not have a prominent MS, we were not able to decisively settle on any one $dA_V$ as MATCH favored models with the highest possible $dA_V$, which maximally spread out the helium burners (HeBs) and thereby gave the minimal fit parameter. Nevertheless, given that \citet{Rao2025} obtained a $dA_V$ of 0.8 for the Southern Arcs, we argue that the true $dA_V$ for ACS Fields 2 and 3 must lie somewhere between 0.0 and 0.8 and choose a fiducial value $dA_V$ of 0.4 for displaying their SFHs and model Hess diagrams. Further, we test for variations in $dA_V$ in the range of $0.0-0.8$ and find that this does not substantially affect the qualitative features of the SFHs on which we base our discussion.

\begin{figure*}
    \centering
    \adjustbox{max height=\dimexpr\textheight-6\baselineskip\relax}{%
        \includegraphics{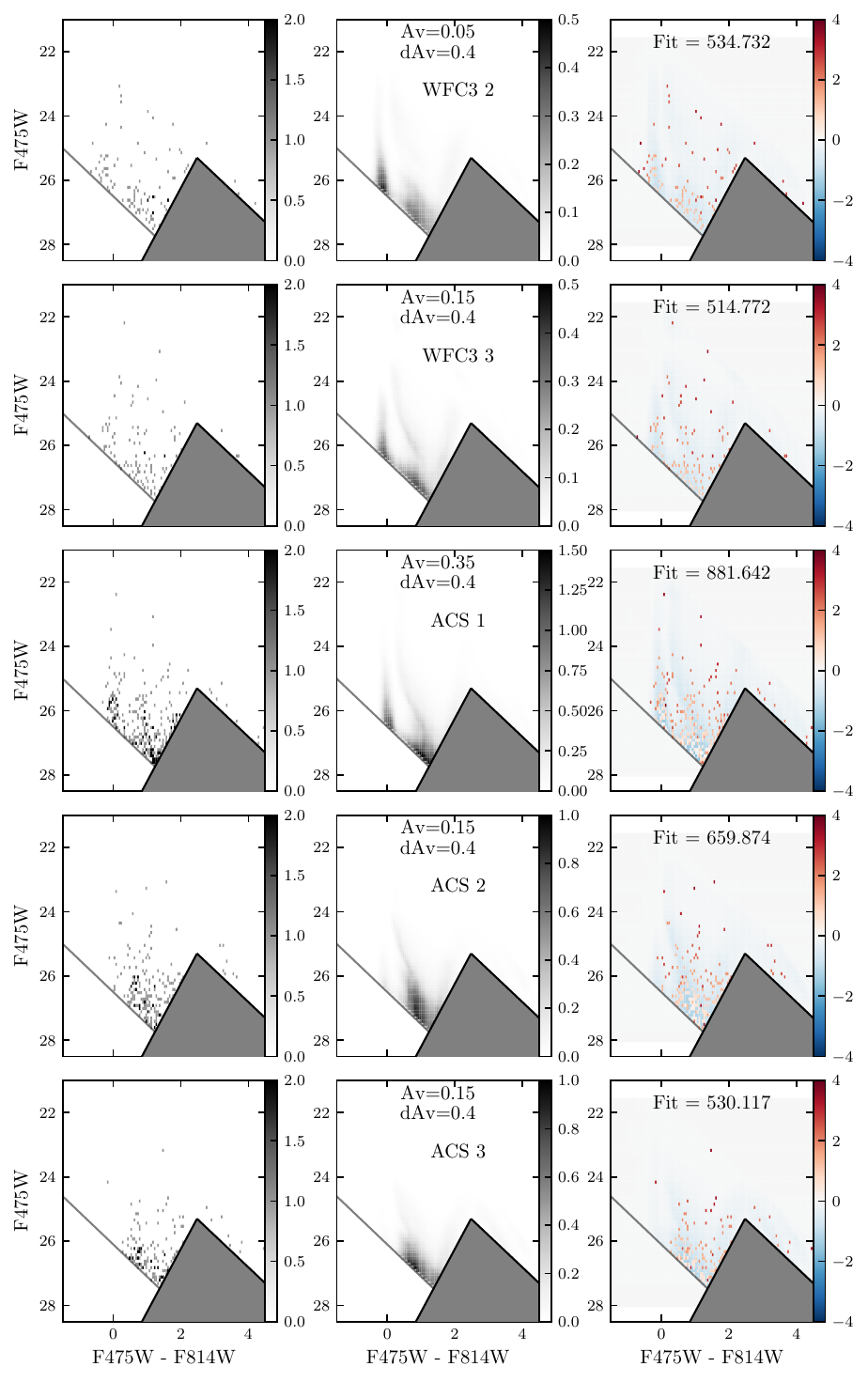}
    }
    
    \caption{For all the Deep Halo Fields, (\textbf{Left}) Observed Hess diagram used by \texttt{MATCH} that balances number statistics on the CMD with time resolution in the SFH; (\textbf{Center}) best-fit model Hess diagram using PADUA stellar evolution models with complete AGB tracks \citep{Marigo2008, Girardi2010}; (\textbf{Right}) residual significance Hess diagram. The color bars in the left and center panels represent the number of stars per CMD bin. In the right panel, the scaling reflects the significance of each pixel in the residual relative to the standard deviation of a Poisson distribution. The RGB mask used is shown in gray in all panels along with the 50\% completeness limit for each field.}
    \label{fig:model-cmds}
\end{figure*}

\begin{figure*}
    \centering
    \adjustbox{max height=\dimexpr\textheight-3\baselineskip\relax}{%
        \includegraphics{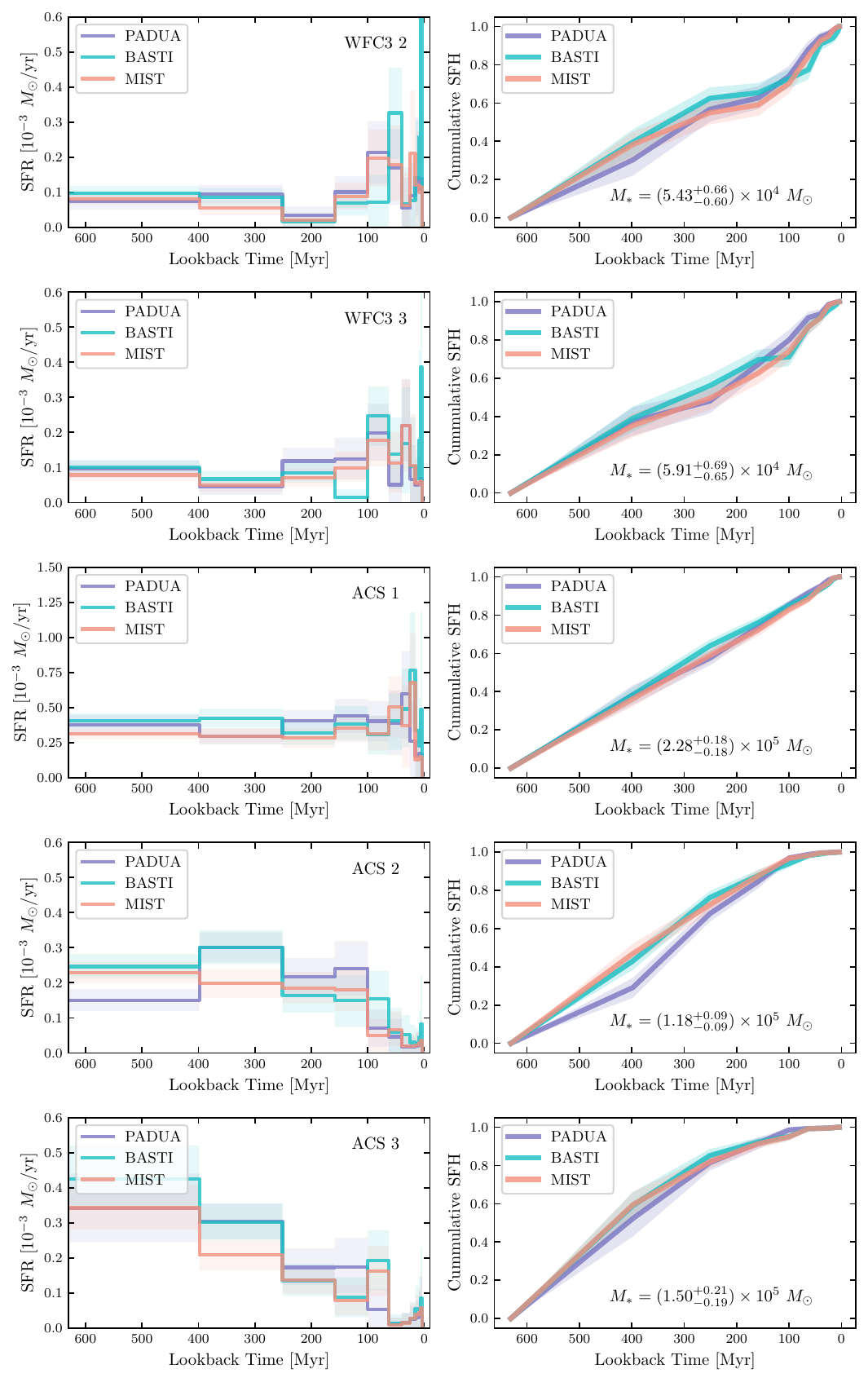}
    }
    \caption{Best fit SFHs for each of the Deep Halo Fields using PADUA, MIST, and BASTI isochrones. Error envelopes represent the 68\% confidence interval of random uncertainties. (\textbf{Left}) Absolute star formation rate (SFR) as a function of lookback time. (\textbf{Right}) Cumulative fraction of stars formed as a function of lookback time. The total stellar mass formed in the last $\sim630\,$Myr according to the PADUA models is also shown.}
    \label{fig:sfh}
\end{figure*}

\subsection{Deep Halo Fields: SFH Results}
In this section, we present the model Hess diagrams and star formation histories of the Deep Halo Fields.

We show the quality of our Hess diagram fits for all the Deep Halo Fields using PADUA models in Figure \ref{fig:model-cmds}. We prefer to show the PADUA models as they consistently give marginally lower residual fit parameters compared to the BASTI or MIST models. The scaling of the residual Hess diagram reflects the pixel-wise deviation of the observed Hess diagram from the model relative to the standard deviation of a Poisson distribution. Overall, we find that the models reproduce distinct features of the observed Hess diagrams such as the horizontal branch, main-sequence turn-off, and main sequence (for those fields that have a distinct MS), matching the data well in terms of densities of stars, luminosities, colors, and scatter.

The SFR  of the Deep Halo Fields as a function of lookback time and cumulative SFH--- fraction of total stellar mass formed prior to a given epoch, are shown in Figure \ref{fig:sfh}. We will mainly interpret the cumulative SFHs as they minimize many of the issues that affect interpreting absolute SFHs such as interpreting covariant SFRs in adjacent time bins and defining appropriate time resolutions \citep[e.g.,][]{H&Z2001, Dolphin2002, McQuinn2010b}. In WFC3 Field 2, the star formation appears roughly constant from $\sim630\,$Myr to $\sim250\,$Myr, forming $\sim60\%$ of the ``young" stellar mass. Following a brief lull, the star formation picks up again at $\sim150\,$Myr in a burst that produces the remaining $\sim40\%$ of the stellar mass. The SFH of WFC3 Field 3 is a ``smoothed out" version of WFC3 Field 2. The star formation appears constant across the $\sim630\,$Myr with signatures of a slight enhancement $\lesssim200\,$Myr. The SFH of ACS Field 1, which is closer to M82's disk, is mostly constant with slight enhancements seen $\lesssim50\,$Myr. ACS Fields 2 and 3 experienced more star formation $\gtrsim200\,$Myr ago compared to $\lesssim200\,$Myr. ACS Field 3, in fact, shows a steadily declining SFR up to $\sim60\,$Myr ago, with virtually no recent star formation. 

The total stellar mass formed in the last $\sim630\,$Myr in each field is detailed in Figure \ref{fig:sfh}. The star formation in the Southern Arcs is $5.3\pm0.3\times10^5\,M_{\odot}$ \citep[]{Rao2025}. Barring the Southern Arcs, if we assume that the remaining Deep Halo Fields cover $\sim50\%$ of the M82 Tail, we estimate the total star formation in the Southern Arcs and M82 Tail in the last $\sim630\,$Myr to be $\sim2\times10^6\,M_{\odot}$.

\section{Young Stars in the ANGST footprint}\label{sec:ANGST}
In this Section, we detail the photometry for the ANGST dataset used to map the young halo stars closer to the M82 disk.

\begin{figure*}
    \centering
    \includegraphics[width=\linewidth]{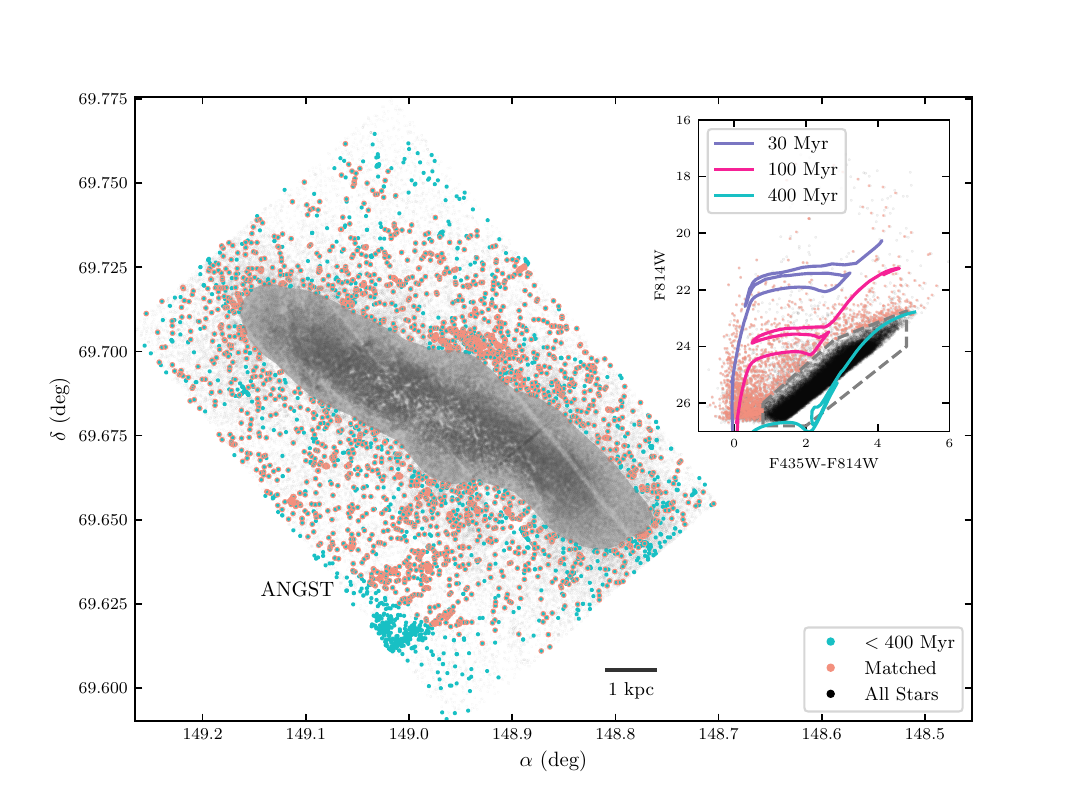}
    \caption{ A map $\lesssim400\,$Myr stars in the ANGST footprint (cyan) with cross-matches in the infrared from the Cibola (NIRCam) survey (salmon). The central disk has been masked out due to crowding and poor photometry. (\textbf{Inset}) CMD of ANGST GST sources in the outer disk and stellar halo with PARSEC isochrones overplotted. The dashed-gray region indicates the CMD mask used to exclude blends. Matched $\lesssim400\,$Myr stars are marked in salmon, and isochrones of various ages are overplotted.
}
    \label{fig:m82-angst-nircam}
\end{figure*}

\subsection{ANGST: Archival Dataset}
HST observations of M82 for the ANGST survey were obtained in March 2006 by \citet{Mutchler2007} and the Hubble Heritage Team (Program 10776) using the Advanced Camera for Surveys’ Wide Field Channel (ACS/WFC). The data include imaging in the F435W, F555W, F814W, and F658N ($H\alpha$) filters across six pointings arranged in a 3×2 mosaic, covering M82 out to $\sim3.5\,$kpc from its disk. Photometry in the F435W, F555W, and F814W bands was produced as part of the ANGST survey and described in detail by \citet{Dalcanton2009}. We selected GSTs from this catalog following the same criteria used for the Deep Halo Fields: \texttt{SNR} $> 4$, \texttt{SHARP}$^2 < 0.2$, and \texttt{CROWD} $< 0.25$.

The ANGST photometry is significantly shallower than the Deep Halo Fields largely due to crowding and dust extinction ($A_V \sim 1$–4; \citealt{Hutton2015}) limitations (see ANGST F435W-F814W vs. F814W CMD, Figure \ref{fig:m82-angst-nircam}). Large portions of the M82 disk appear to have unreliable photometry owing to ``blends" from multiple crowded sources, which are not removed even after incorporating strict quality cuts. We thus generate a spatial density map of the stellar catalog using a kernel density estimator (using \texttt{scipy}'s \texttt{gaussian\_kde}) and mask out the central dense regions at density levels $>20$. Even after employing the disk mask, there exists a significant amount of blends at F435W$>27$, which approaches the 50\% completeness limits of F435W at 27.3 \citep{Dalcanton2009}. To remove these blends and the ancient RGB stars, we incorporated an additional mask in F435W-F814W vs. F814W CMD space. This leaves us with 3410 MS and HeB stars of ages $\lesssim400\,$Myr that live in the outer disk and stellar halo of M82. We will henceforth refer to these stars as ``young ANGST stars".

The young ANGST stars are ubiquitous on either side of the M82 disk (Figure \ref{fig:m82-mosaic}). Notably, there are hints of structures in these young stars such as overdensities immediately to the north and south of the disk, coincident with regions where the outflow emerges. There are also what appear to be filaments of stars to the south and extensions of the Southern Arcs into the ANGST footprint. However, given the sparse distribution of these stars, we will leave the interpretations of these structures to future deeper surveys. 

We may however, estimate the total mass of stars formed in M82's stellar halo traced by the number of young ANGST stars. To do this, we generated a mock stellar population using \texttt{MATCH}'s \texttt{fake} functionality. We assumed a constant SFR of 0.01 $M_{\odot}/\mathrm{yr}$, used logarithmic age bins from 6.6-8.8 (approximately $4-400\,$Myr) with a spacing of 0.2 dex, assumed a constant solar metallically with a spread of 0.2 dex, assumed an $A_V$ of 0.4 and a $dA_V$ of 0.8, used Kroupa's IMF, and incorporated the ANGST magnitude and completeness limits. With a constant SFR, the IMF ensures that the number of stars in our region of interest scales linearly with SFR. We thus count the number of stars that pass the CMD mask described above and compare that to the number of young halo stars to obtain a SFR estimate. We find the halo SFR in the last $\lesssim400\,$Myr to be $\sim4\times10^{-3}\,M_{\odot}/\mathrm{yr}$, forming $\sim10^{6}\,M_{\odot}$ of stars.

\subsection{Validating ANGST stars with JWST}
Given that so many of the stars of interest in ANGST lie within one magnitude of the F435W completeness limit, we felt it necessary to check these stars against a deeper resolved star dataset to ensure they weren't photometric errors. We thus attempt to validate the photometry for these young halo stars using deep JWST NIRCam photometry of M82 and its halo from the Cibola survey (GO5145; PI: Dr. Adam Smercina; \textit{priv. comm.}). The Cibola survey covers roughly the same area of M82 as that covered by ANGST.

GST sources in the ANGST and Cibola photometery catalogs were matched based on their sky positions (using \texttt{astropy}'s \texttt{SkyCoord})--- match candidates were required to be within a 0.1 arc-second coordinate separation between datasets. We matched only the outer disk and halo stars using the disk mask employed for the ANGST dataset, which yielded 42913 matches. We confirmed the matches by making cross instrument color-color plots (F435W-F814W vs F115W-F200W) to verify that matches lie along the stellar locus within photometric uncertainties. There were, however, some spurious matches that did not lie along the stellar locus. These spurious sources were largely a result of photometric errors in the ANGST dataset arising from blends--- using stars well above the F435W 50\% completeness limit (F435W$<27$) and limiting matches to stars to the left of the $\lesssim 400\,$Myr isochrone in Cibola's F115W-F200W vs. F200W CMD eliminated these outliers from the color-color plots. We found that 86\% of the young stars in the ANGST footprint have matches in Cibola. This tells us that apart from a few spurious sources, most of the blue stars in ANGST are also blue in the infrared, increasing our confidence in the existence of these young stars throughout M82's halo. In other words, ANGST F435W was indeed deep enough to detect most of the $\lesssim 400\,$Myr stars in the halo regions with reasonable purity. 

Factoring in the contamination fraction and accounting for the completeness of the photometry in our regions of interest in both real and CMD space ($97\%$ from ASTs; \citealt{Dalcanton2009}), the average SFR in the stellar halo in the last  $\lesssim400\,$Myr would be $\sim3\times10^{-3}\,M_{\odot}/\mathrm{yr}$, forming $\sim10^{6}\,M_{\odot}$ of stars. Extrapolating to $630\,$Myr, the star formation is $\sim2\times10^{6}\,M_{\odot}$.

\section{Physical Origins of Young Halo Stars in M82}\label{sec:discussion}
In Sections \ref{sec:subaru}, \ref{sec:hst-deep}, and \ref{sec:ANGST}, we showed that stars with ages $\lesssim630\,$Myr are distributed widely throughout M82’s CGM. These stars surround the stellar disk across the full ANGST footprint, form distinct kiloparsec-scale arc-like structures--- the Southern Arcs--- and extend to projected distances of $\sim20\,$kpc eastward in the form of the M82 Tail. Along the Tail, the SFHs derived in Section \ref{sec:hst-deep} show signs of enhanced recent star formation at smaller galactocentric distances. Together, these results point to ubiquitous, \textit{in-situ} star formation in M82’s halo.

What physical processes give rise to these young halo populations, and what governs their striking spatial distribution? To address these questions, we now compile the observational evidence presented above and examine the physical ingredients capable of producing widespread halo star formation. We consider a range of plausible scenarios and, where useful, augment the discussion with illustrative visualizations and order-of-magnitude estimates.

\begin{figure*}
    \centering
    \includegraphics[width=\linewidth]{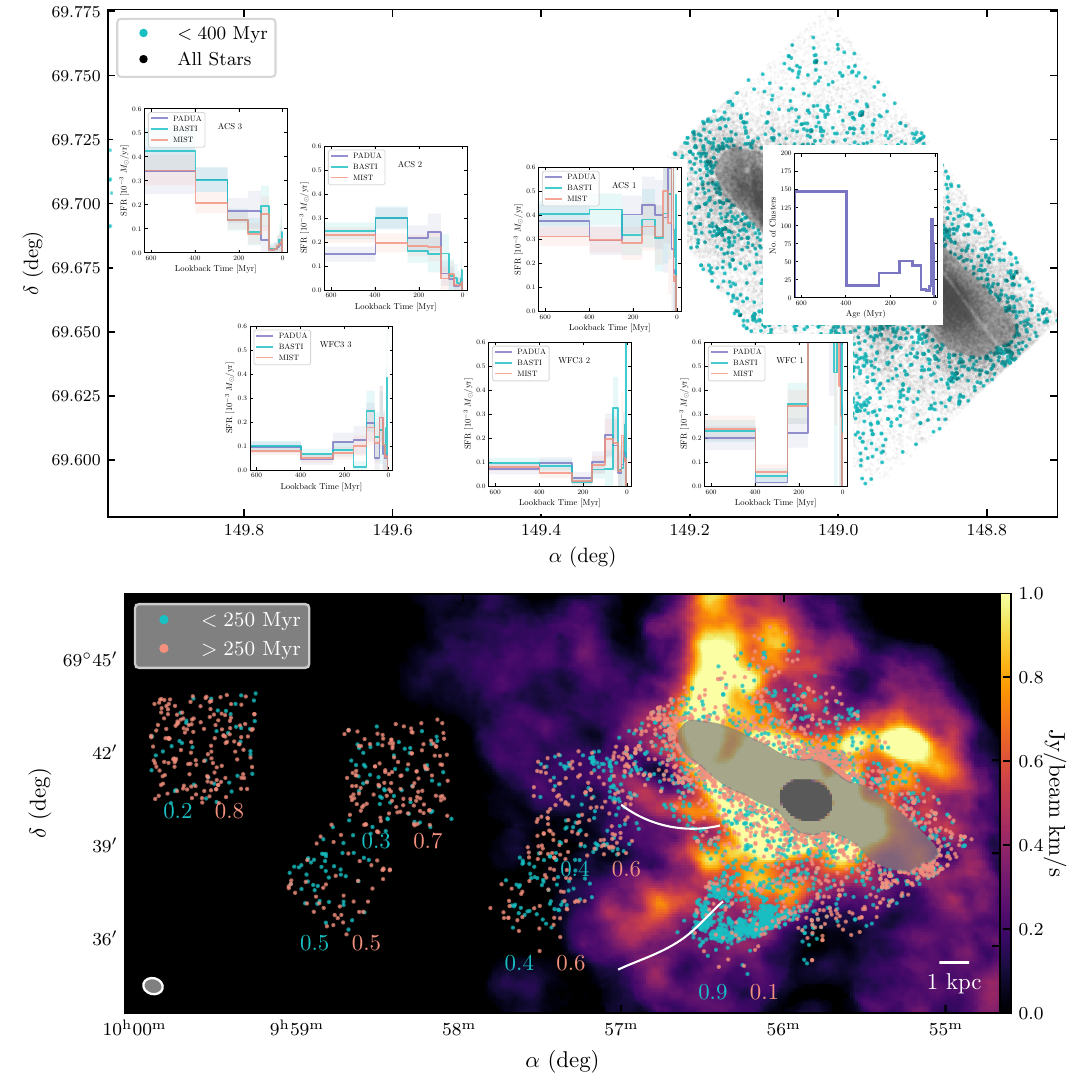}
    \caption{(\textbf{Top}) The SFHs of the Deep Halo Fields visualized with the SFH of the M82 disk. Each panel shows the SFR as a function of time with identical $y$-axis scales and time bins. The SFH of the Southern Arcs was re-derived using \texttt{MATCH} with the same parameters as \citet{Rao2025}, but adopting the coarser time bins used in this work. The M82 disk star cluster age distributions have been reproduced from \citet{Li2015}. (\textbf{Bottom}) The distribution of neutral hydrogen \citep[HI;][]{deBlok2018} overlaid with the $\lesssim630\,$Myr stars from ANGST and Deep Halo Fields. We color-code stars $\lesssim 250\,$Myr in cyan and stars $\gtrsim250\,$Myr in salmon, selecting the helium burners from the CMDs of each field using PARSEC isochrones. Below each Deep Halo Field, we show the fraction of stellar mass formed in each age range with their corresponding color. Spurs of HI trailed by the stars in the Deep Halo Fields have been highlighted by white arcs.}
    \label{fig:sfh-viz}
\end{figure*}

\subsection{Outflow-driven Star Formation}
The first possibility that we consider for the origin of the $\lesssim630\,$Myr stars throughout M82's stellar halo is star formation induced by M82's starburst outflow. Given that the Southern Arcs were likely born out of the outflow \citep{Rao2025}, it is possible that the young stars that populate the halo elsewhere share a similar origin. We see evidence in favor of this picture in the SFHs of the Deep Halo Fields when compared to the SFH of the disk--- traced by the age distribution of star clusters (Figure \ref{fig:sfh-viz}). 

Star clusters in the M82 disk have been identified, cataloged, and studied through a variety of optical and near-infrared observations using several facilities including HST and JWST \citep[e.g.,][]{McCrady2003, Mayya2008, Lim2013, Li2015, Levy2024}. The catalog of 1357 star clusters identified through JWST's NIRCam by \citet{Levy2024} is the most complete. However, these clusters have not been aged. Ages, have only been derived for a smaller sample of 846 star clusters identified through integrated light, multi-band HST images \citep{Lim2013, Li2015}. We show the age distribution of these clusters in Figure \ref{fig:sfh-viz} after binning the ages of these clusters to be consistent with the time bins used for our SFH fits for the Deep Halo Fields. In our period of interest, the age distribution reveals a period of intense star cluster formation from $\sim630\,$Myr to $\sim400\,$Myr. We also see a period of heightened star formation beginning $\sim200\,$Myr ago and lasting until $\sim70\,$Myr. Based on these star cluster ages, we will refer to stars $\gtrsim250\,$Myr and stars $\lesssim250\,$Myr for simplicity.

Stars corresponding to each of these two ``epochs" of heightened star formation in the disk are found throughout the Deep Halo Fields and in the halo immediately around M82 (see lower panel, Figure \ref{fig:sfh-viz}). In particular, WFC3 Fields 2 and 3 have a higher fraction of $\lesssim250\,$Myr stars compared to the other fields, with a period of heightened star formation $\sim100\,$Myr ago--- similar to the Southern Arcs. ACS Fields 2 and 3 on the other hand, have a higher fraction of stars formed $\gtrsim250\,$Myr. In ANGST, despite being limited in sensitivity to star ages $\lesssim 400\,$Myr, stars corresponding to both the $\gtrsim250\,$Myr and $\lesssim250\,$ epochs are distributed throughout the footprint. The case, then, for outflow induced star formation is not straightforward. If the periods of heightened star cluster formation in the disk correspond to the launching of galactic-scale starburst outflows, stars formed at varying levels at different locations in the Deep Halo Fields due to the outflow's influence. 

\citet{Rao2025} proposed two mechanisms for the formation of the Southern Arcs, driven by a $\sim100\,$Myr starburst event in the M82 disk. The first involved a shock front produced by the hot gas outflow punching through the surrounding CGM, compressing CGM clouds or pieces of tidal debris, inducing star formation. The second involved stars forming directly within Jeans-unstable cool $\lesssim10^4\,$K clouds that were either launched from the disk or that were precipitated out of the mixing layers within the outflow. This second mechanism involves stars inheriting the velocity of the out-flowing cool clouds and traveling along ballistic trajectories. 

If stars in the M82 Tail formed through this second mechanism, the trajectories would not align with the Eastward orientation of the tail since M82's southern outflow is pointed southeast. The mechanism of shock-induced star formation on the other hand, may be favorable. Comparing Chandra X-Ray Observatory soft X-ray maps of M82's outflow with three dimensional simulations of the starburst outflow of \citet{Cooper2008}, \citet{Rao2025} point out that the influence of shocks from the outflow may be much more widespread compared to the apparent bi-conical morphology of the outflow as the soft X-ray emitting shock interface between the outflow and the ambient medium takes on a nearly spherical geometry. These shocks, unconstrained by the direction of the outflow, could have overrun gas clouds of varying densities and masses, distributed to the east in M82's CGM, inducing star formation in these clouds.

\subsection{The Role of Stripped Gas and Streamers}
Within the framework of halo star-formation driven by outflow shocks, the distribution of star formation is governed by the distribution of pre-existing gas clouds in the CGM. In the vicinity of M82, these gas clouds may be gas stripped from the M82 disk due to tidal interactions with M81 \citep[e.g., the Northeast and Southwest Spurs;][]{Yun1993} or they may be linked to the cool material that sheaths the bi-cone of M82's hot wind \citep{Leroy2015, Martini2018}.

\begin{figure}
    \centering
    \includegraphics[width=\linewidth]{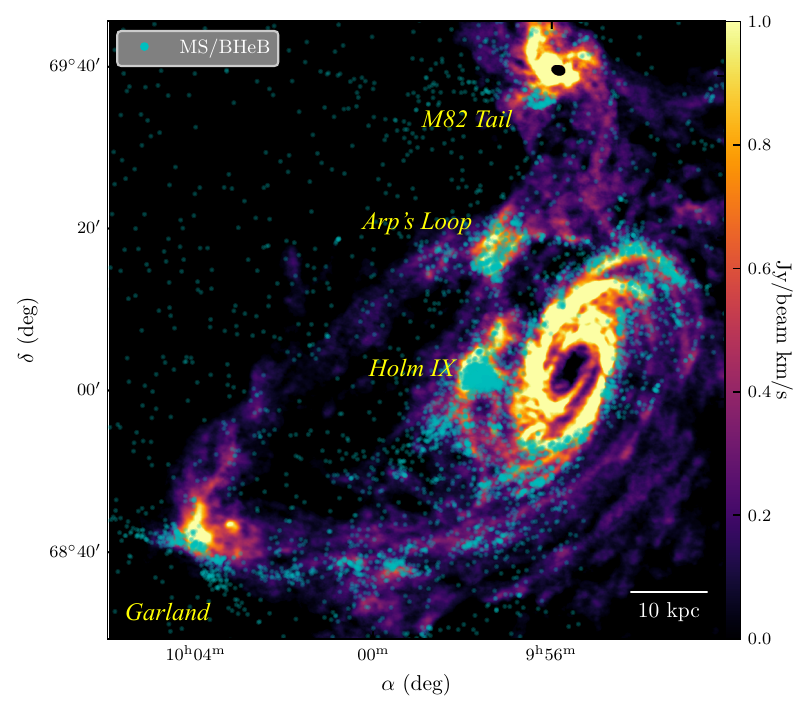}
    \caption{A map of the neutral hydrogen \citep[e.g.,][]{deBlok2018} in the M81 Group with young ($\lesssim400\,$Myr) MS and BHeB Subaru stars over-plotted. Prominent star-forming tidal features--- Arp's Loop, Holmberg IX, and the Garland--- have been labeled along with the M82 Tail. Young stars in the M82 Tail and Garland are offset from nearby HI by several kiloparsecs unlike Holmberg IX and Arp's Loop.}
    \label{fig:m81-h1}
\end{figure}

The panoramic view of young stars surrounding M82 (Figure \ref{fig:m81-subaru}) shows that, more than a few kiloparsecs from M82, young stars live predominantly to the east in the M82 Tail. This suggests that the gas clouds from which these young stars were born were preferentially distributed to the east of M82. Despite the fact that we do not see any particular HI (or any other gas for that matter) currently distributed along the Tail (Figure \ref{fig:sfh-viz}, \ref{fig:m81-h1}), it is plausible that gas clouds existed to M82's east in the past, which dissipated after forming stars. In fact if we trace the Tail towards M82's disk, we do come across spurs in HI that extend more or less in the direction of the Tail (see lower panel, Figure \ref{fig:sfh-viz}). However, due to projection effects, it may be that these spurs are not associated with the Tail at all.

Star formation in stripped gas far from the main bodies of galaxies has been extensively documented both within the M81 Group \citep[e.g.,]{Makarova2002, S&M2001, Okamoto2015} and in other interacting galaxies such as those in the Virgo cluster \citep[e.g,][]{Kenney2014,Boselli2018}. Holmberg IX, Arp's Loop, and the Garland are some of the best examples within the M81 Group. HST imaging of these features have identified dominant star forming episodes $\lesssim300\,$Myr for the Garland and $\lesssim100\,$Myr for Holmberg IX and Arp's Loop \citep{Makarova2002, Weisz2008}. Holmberg IX and Arp's Loop both have an underlying distribution of HI, while lacking an underlying RGB population. The Garland also lacks an underlying RGB population, but, the HI is offset from the young stars by several kiloparsecs (Figure \ref{fig:m81-h1}). The M82 Tail is similar to all three of these ``tidal dwarfs" in terms of having stars of similar ages and no associated RGB stars. However, it differs from Holmberg IX and Arp's Loop since its association with HI features are tentative at best, more similar to the Garland. This offset between the stars and HI in both the Garland and the M82 Tail could be a feature of the longer timescales of star formation--- providing sufficient time for the stars formed to decouple from or disperse their natal gas.

The above comparisons suggest that stars in Holmberg IX, Arp's Loop, Garland, and M82 Tail were potentially born out of clouds of stripped gas, but the M82 Tail is unique in the sense that, overall, the star formation histories appear similar to that of the M82 disk and the Southern Arcs. This could be indicative of gas being a necessary condition for star formation, but not a sufficient one. In the case of the tidal dwarfs, gas densities may have been sufficiently high enough to induce spontaneous star formation, but in the case of the M82 Tail, an external trigger such as shocks from M82's outflow may have been necessary, given its correlation with star formation in the M82 disk.

\subsection{The Ram Pressure Stripping Scenario}
While the halo star formation in the M81 Group is mostly in tidally stripped gas, halo star formation in Virgo cluster galaxies occurs in ram pressure stripped gas. This suggests an interesting possible third origin for cold gas around M82 in addition to tidal stripping and outflow launching. The cold gas in M82's CGM could be to ram pressure stripping of the gas disk of M82 or the stripping of material launched into the CGM by M82's mass-loaded outflow, as M82 moves through the ambient medium of the M81 Group. If the gas from which the M82 Tail stars were born were indeed from ram-pressure stripping, it would suggest that M82 is moving west.

This westward motion runs counter to N-body simulations of the M81 Group, which suggest that M82 is moving northeast \citep{Yun1999}. However, these simulations were primarily constrained by the HI morphology of the group from data available at that time \citep{Yun1994} and did not incorporate the two decades of additional observational evidence since gathered--- such as resolved stellar population maps spanning the group \citep[e.g., this work;][]{Okamoto2019, Smercina2020}) and deeper HI surveys with improved kinematic measurements \citep{deBlok2018, Martini2018}. 

Apart from the morphology of the Tail, a key piece of evidence in favor of M82's westward motion is the ages of stars in the M82 Tail. The fraction of stellar mass formed $\gtrsim 250\,$Myr decreases westwards and southwards along the trail, while the fraction of stellar mass formed $\lesssim250\,$Myr increases (Figure \ref{fig:sfh-viz}). This would be consistent with the picture of gas farther from M82 being stripped at earlier times compared to gas closer. If this stripped gas formed stars under the influence of outflow shocks, we would see predominantly older stars farther away from M82. The balance of stars in each of these age ranges indicates a dominant star formation gradient. However, the fact that we see stars belonging to both these age ranges in all locations is puzzling. It could be that the stripped gas farther away continued forming stars at low levels following the dominant star formation episode and there could have been some dynamical mixing of these older stars that could have placed them closer to M82 in projection. This model for explaining age gradients in stars formed in ram pressure stripped material has been explored for example in the Virgo Cluster galaxy IC3418 \citep{Kenney2014}. 


\subsubsection{Proper Motion \& Projection Effects}
In the ram pressure stripping scenario, we may also place a constraint on the proper motion speed of M82 relative to the proper motion of the M81 Group. The oldest epoch of star formation that the SFHs are sensitive to is $\sim400-630\,$Myr in ACS Field 3, which is $\sim20\,$kpc from M82 in projection. Assuming that the stripped gas comes to a standstill with respect to the ambient medium of the M81 Group, these numbers would imply a proper motion speed of $v_{\mathrm{M82,sky}} \sim 30-50\,\mathrm{km\,s^{-1}}$. Additionally, M82 has a redshift velocity of $308\pm5\,\mathrm{km\,s^{-1}}$ with respect to M81 \citep{Speights2012, vanderTak2008}. Further assuming that the ambient medium of the M81 Group moves with M81, we may thus derive the true speed of M82 relative to M81 to be $v_{\mathrm{M82-M81}}\sim309-312\,\mathrm{km\,s^{-1}}$, suggesting most of M82's motion is in the direction into and normal to the sky plane.

\begin{figure*}
    \centering
    \includegraphics[width=\linewidth]{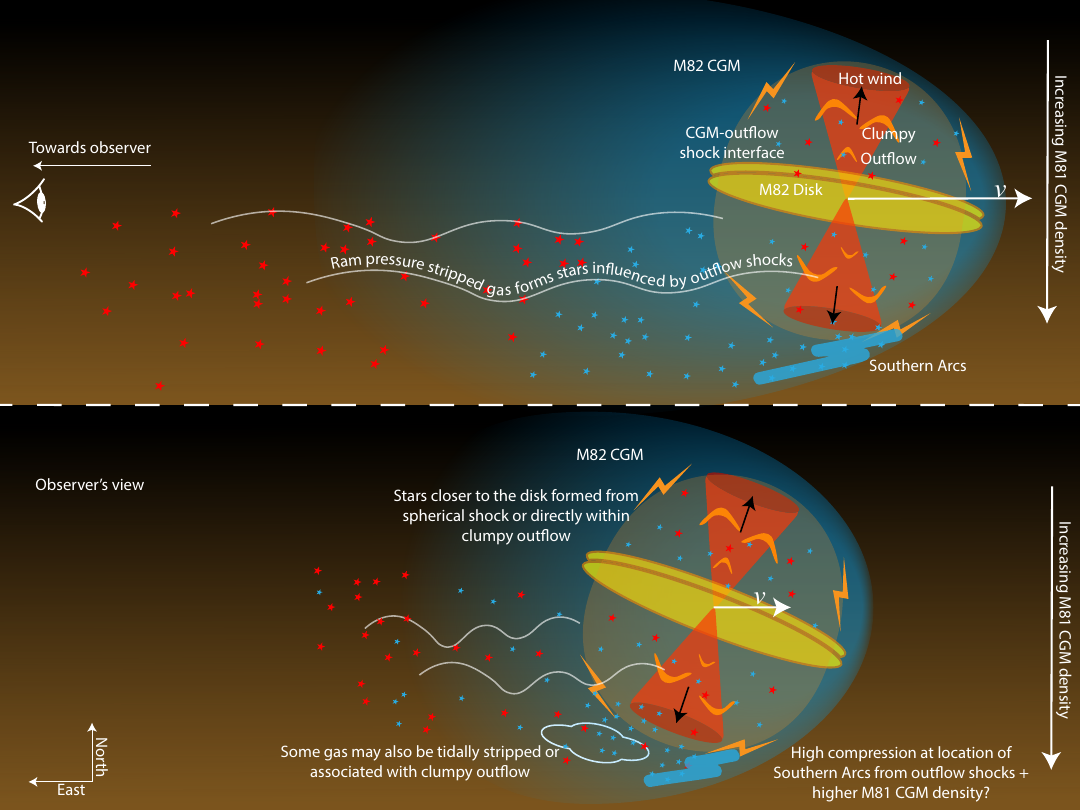}
    \caption{An illustration of the various contributors to star formation in M82's halo. (\textbf{Top}) M82 viewed perpendicular to its line of sight to the observer. It is moving at a velocity of $269\,\mathrm{km\,s}^{-1}$ away from the observer. (\textbf{Bottom}) M82 from an observer's perspective. The galaxy is probably moving West relative to M81's CGM. Stars in the halo close to the M82 disk may have formed directly in M82's clumpy outflow or in hurled or stripped molecular gas clouds in the CGM. Stars in the M82 Tail may have formed in ram pressure stripped gas and could have been influenced by outflow shocks. The location of the Southern Arcs may have been a favorable location for star formation owing to the proximity of the outflow and compression of M82's CGM from its motion through the ambient medium.}
    \label{fig:m82-motion-illustration}
\end{figure*}

The above velocity considerations have important consequences. If the proper motion of M82 is $\sim40\,\mathrm{km\,s^{-1}}$ to the west and radial motion $308\,\mathrm{km\,s^{-1}}$ into the sky plane, both relative to M81, the true extent of the M82 Tail would be $\sim160\,$kpc instead of the $20\,$kpc viewed in projection. This suggests that the $\lesssim250\,$Myr stars in the M82 Tail may be significantly closer to M82 compared to the $\gtrsim250\,$Myr stars, making the apparent distribution of stars from both populations everywhere in the Tail an illusion from projection (see Figure \ref{fig:m82-motion-illustration}).

\subsubsection{Ram Pressure Estimates}
Here, we attempt to quantify the ram pressure acting on M82's CGM arising from its motion through M81's ambient medium. We assume that the stripping occurs roughly at the location of the Southern Arcs and M82 Tail $\sim10\,$kpc from M82 and $\sim35\,$kpc from M81 due to the hot gas in M81's CGM. To calculate the hot gas density $\sim 35\,$kpc from M81,
we assume that the hot gas in M81's CGM follows the same density profile as that of the Milky Way, represented by a spherical $\beta$-model. The $\beta$-model is defined as
\begin{equation}
    n(r) = n_0(1+(r/r_c)^2)^{-3\beta/2}
\end{equation}
where $n_0$ is the central number density, $r_c$ is the core radius, and $\beta$ defines the slope of the profile at large radii. From OVIII emission line measurements, \citet{Miller2015} find best-fit parameters of $n_0r_c^{3\beta}= 1.35\pm0.24\,\mathrm{cm}^{-3}\,\mathrm{kpc}^{3\beta}$ and $\beta= 0.50\pm0.03$. Since the typical core radii $\lesssim5\,$kpc, we adopt the modified $\beta$-model in the limit $r>>r_c$:
\begin{equation}
    n(r)\approx\frac{n_0r_c^{3\beta}}{r^{3\beta}}
\end{equation}
Assuming complete ionization and a metallicity of $Z=0.3\,Z_\odot$, we find that the hot gas $\sim35\,$kpc from M81 has a mass density of $\sim7\times10^{-30}\,\mathrm{g}\,\mathrm{cm}^{-3}$.

The ram pressure is defined to be $P_{\mathrm{ram}}\sim\rho v^2$ \citep{Gunn&Gott1972}. In this case, $\rho$ is the density of the gas in the M81 CGM and $v$ is the relative velocity of M82 relative to M81's CGM. Using $v\sim310\,\mathrm{km\,s}^{-1}$, we find $P_{\mathrm{ram}}\sim10^{-14}\,\mathrm{dyn\,cm^{-2}}$. For material to be stripped from M82's CGM, the ram pressure exerted on the gas must exceed the gravitational restoring force per unit area. 

Following the work of \citet{McCarthy2008}, we assume the ram pressure stripping of a spherically symmetric distribution of gas. The stripping condition with the gravitational restoring force can be expressed as
\begin{equation}
    P_{\mathrm{ram}}> \Sigma_{\mathrm{gas}}(R).g(R)
\end{equation}
where $\Sigma_{\mathrm{gas}}(R)$ is the projected surface density of gas in the sphere along the direction of ram pressure and $g(R)\sim GM_{\mathrm{gal}}(R)/R^2$ is the gravitational acceleration from M82 ($R$ in this case is the distance from M82). The condition yields an upper limit on the projected surface density of gas in M82's CGM that can be ram pressure stripped. Since, most of M82's motion is along the line-of-sight, this maximum surface density yields an upper limit on the observed column density of HI. We use the NFW dark matter halo mass model \citep{NFW1996} for M82 detailed in \citet{Oehm2017} and \citet{Martini2018} to calculate the restoring acceleration at $R\sim10\,$kpc and assume a neutral, solar metallicity for the stripped gas. We thus find that a ram pressure of $\sim10^{-14}\,\mathrm{dyn\,cm^{-2}}$ can strip HI of column densities at most $10^{18}\,\mathrm{cm}^{-2}$ from M82's CGM.

Present-day neutral hydrogen at the location of the Southern Arcs ranges from $3-6\times10^{20}\, \mathrm{cm}^{-2}$. In addition, cool clouds detected in M82's outflow range between $10^{20}-10^{21}\,\mathrm{cm}^{-2}$ \citep{Lopez2025}. This suggests that it is likely that only the diffuse gas in M82's CGM may be stripped due to ram pressure from M81's hot gas and not the dense/clumpy gas in M82's disk or mass-loaded outflow. Since the minimum column density required for star formation typically ranges from $3-10\times10^{20}\, \mathrm{cm}^{-2}$ \citep{Schaye2004}, it is likely that outflow shocks may well have been necessary to allow some fraction of the (lower density) gas that could be ram-pressure stripped to be compressed to the densities required for star formation. 

\subsubsection{A Unique Flavor of Ram Pressure Stripping}
The objective of the above calculations is not to re-establish the existence of star formation in ram-pressure–stripped gas, which is already well documented. Star formation in stripped material has been observed in extreme cluster environments, most notably in jellyfish galaxies such as ESO 137-001 \citep[e.g.,][]{Sun2007}, as well as in Virgo Cluster systems including IC 3418 and NGC 4254 \citep[e.g.,][]{Hester2010, Kenney2014, Boselli2018}. These galaxies, however, are moving through the dense intracluster medium characteristic of massive clusters. M82, by contrast, resides in a galaxy group, where the ambient medium is substantially less dense. The M82 Tail may therefore represent a lower-pressure, and consequently weaker, manifestation of the same extraplanar star formation phenomenon seen in cluster galaxies.

What sets M82 apart is the presence of its powerful starburst-driven outflow, which gives a unique flavor to this ram-pressure associated star formation. Repeated starburst episodes may replenish M82’s CGM with gas that is subsequently susceptible to ram-pressure stripping. In addition, shocks driven into the CGM by the outflow may further compress this stripped material, aiding star formation in an environment where ram pressure alone would otherwise be marginal.

\subsection{Disk Heating and Stellar Splash?}
So far, we have only explored gaseous origins for the distribution of young stars in M82's halo. However, potential non-gaseous, dynamical scenarios exist \citep[see for e.g.,][]{Dorman2013, Johnston2017}. For example, minor accretion events and mergers can dynamically eject disk stars into the stellar halo \citep[e.g,][]{Zolotov2009, Purcell2010, Font2011}. Tidal forces arising from close encounters can strip stars from the main galaxy disk resulting in tidal tails and bridges \citep[e.g.,][]{Toomre&Toomre1972, Gnedin2003}. Dynamical heating by molecular clouds, spiral arms, and star formation activity can cause stars to leave the disk plane and populate the stellar halo \citep[e.g.,][]{Spitzer&Schwarzschild1951, Barbanis&Woltjer1967, Kroupa2002}. A common feature of all such mechanisms is that they do not discriminate between stars of different ages--- dynamical heating and tidal forces would kick-out/pull a mix of old and young stars into the stellar halo. 

In the case of M82, the $\lesssim630\,$Myr stars in the halo are sparsely distributed. The Southern Arcs are an exception, having a clear over-density of $\lesssim100\,$Myr stars with no underlying ancient RGB star population \citep{S&S2024, Rao2025}. In the Subaru map (Figure \ref{fig:m81-subaru}), there is no obvious RGB star over-density underlying the M82 Tail, ruling out a dynamical origin. However, apart from the extensions of the Southern Arcs in the ANGST footprint, the sparse distribution of the young stars in ANGST makes it impossible to say for sure whether these stars are associated with an underlying ancient RGB population as RGB stars exist everywhere in M82's stellar halo. Thus, it is possible that some of these young stars distributed within $\sim3.5\,$kpc of the M82 disk plane could have dynamical origins.

In general, the vertical distribution of stars in edge-on disk galaxies has an age dependence due to disk heating--- younger stars are found closer to the disk compared to older stars. In addition, the scale height of the young component in low mass disk galaxies is larger than that of the Milky Way, as stars in low mass galaxies form in thicker disks \citep{Seth2005}. In their sample of nearby, edge-on, low mass disk galaxies, \citet{Seth2005} detect stars up to $3.5\,$kpc above the disk plane similar to M82. However, within the confines of our rough population selection, we do not see a difference in the distributions of the $\gtrsim250\,$Myr and $\lesssim250\,$Myr populations on either side of the disk (Figure \ref{fig:sfh-viz}). If these stars were indeed thick disk objects, we would expect the younger population to hug the disk plane. These arguments suggest that while the majority of young stars may not be from a thick disk, some stars could be from the thick disk as seen in the \citet{Seth2005} sample \citep[see also][]{Davidge2008b}. 

In addition to disk heating, \citet{Davidge2008b} consider some more dynamical scenarios in M82. They explore the possibility of some of the extraplanar stars being runaway objects--- stars having peculiar motions $\gtrsim30\,\mathrm{km\,s^{-1}}$ directed out of the disk plane due to dynamical processes. They conclude that while it is difficult for $\lesssim100\,$Myr stars to populate the full extent of the extraplanar regions as they would require high peculiar velocities, some fraction of the stars closest to the disk plane may have a runaway origin. A second possibility they consider is tidal stripping. Such a process would pull stars of all ages out of the disk, which is consistent with the mixture of stars of all ages seen in the ANGST footprint. However, they note that within this framework, stars with ages $\lesssim100\,$Myr are hard to explain as these would have to be stripped well after M81 and M82's closest encounter. 

A third interesting possibility is the tidal dissipation or disruption of structures that formed due to the influence of the starburst outflow similar to the Southern Arcs. Assuming that these structures disperse within a few crossing times, a structure formed $\sim10-20\,$kpc from M82 would disperse in $\sim10^8-10^9$ years. If these structures formed at distances $\lesssim10\,$kpc from M82, this could in principle explain the distribution of all the young stars within the ANGST footprint. The crossing argument also explains why the M82 Tail, extending potentially $\sim10-100\,$kpc has not yet dispersed as it would take $\sim10^9 - 10^{10}$ years for such a structure to disperse if it were gravitationally bound to M82.

\section{Synthesis}\label{sec:summary}
In this work, we present the most detailed map to date of young ($\lesssim630\,$Myr) stars in M82’s stellar halo. The following are observational facts:
\begin{itemize}
    \item Stars are ubiquitously distributed on both sides of M82’s disk out to projected distances of $\sim3.5$\,kpc. While this extent represents a modest fraction of the disk radius of a galaxy like M81 \citep[14.7\,kpc;][]{deVaucouleurs1991}, it corresponds to a substantial axial ratio of $\sim51\%$ for M82, given the radius of its optical disk is only 6.9\,kpc \citep{Mayya2009}. Beyond this distance, the distribution of stars becomes asymmetric, with clear detections extending $\sim5\,$kpc to the south in the Southern Arcs and up to $\sim20\,$kpc to the east in the M82 Tail (in projection).
    \item  The total halo star formation in the last $630\,$Myr is $\sim4\times10^6\,M_{\odot}$ ($\sim2\times10^6\,M_\odot$ in the Southern Arcs $+$ M82 Tail and $\sim2\times10^6\,M_\odot$ in ANGST), yielding an average SFR of $\sim7\times10^{-3} M_{\odot}\,\mathrm{yr}^{-1}$. This is not a significant fraction of the $\sim10M_{\odot}\,\mathrm{yr}^{-1}$ SFR that the M82 disk has been experiencing for the last $\sim50\,$Myr \citep{deGrijs2001}.
    \item Star formation progressively becomes older when moving away from M82 along the M82 Tail. Formation histories derived for the stars in the Tail reveal that, overall, their ages are broadly correlated with periods of heightened star cluster formation in the M82 disk, suggesting that M82's outflow was a key ingredient in their formation.
\end{itemize}
We have discussed several possible ingredients and methods to explain this kind of widespread extraplanar star formation in the stellar halo. While it is likely that multiple star forming processes operated in tandem, we attempt to synthesize a physical scenario that highlights important contributing factors (Figure \ref{fig:m82-motion-illustration}).

Our setup involves M82, its outflow, and its CGM moving through the ambient medium of the M81 Group, which arguably increases in density southwards, towards M81. Given the morphology of the Southern Arcs and the M82 Tail, we favor the framework within which stars in the Tail were formed in gas stripped from M82's CGM by the ram pressure of the hot gas present in the M81 Group's ambient medium. Redshift velocities together with the orientation of the Tail and ages of stars in the Tail indicate that the bulk of M82's motion relative to the M81 Group is into the sky plane, away from the observer, with a small component of motion to the west. Back of the envelope estimates suggest that only the diffuse gas in M82's CGM ($N_H \lesssim 10^{18}\,\mathrm{cm}^{-2}$) may be susceptible to ram pressure stripping by the hot gas in M81's CGM. If the density of gas clouds after stripping were sufficiently high, stars may have formed directly within the stripped material. However, shocks driven by M82’s outflow may also have promoted star formation in this gas, particularly given the observed correlation between the stellar ages in the M82 Tail and those in the M82 disk.

The Southern Arcs may have formed in a similar, albeit much more dramatic fashion, with shocks driven by the outflow overrunning clouds of CGM gas stripped by tidal or ram pressure forces. Considering the fact that M81's CGM increases in density southwards, the location of the Southern Arcs may have been an exceptionally favorable location for star formation with strong compression of the gas from outflow shocks combined with higher ram pressure forces from M81's CGM. However, the Southern Arcs may have also formed directly within cool clouds associated with the outflow, launched by a particularly powerful starburst event $\sim100\,$Myr ago. Closer to the M82 disk, both mechanisms--- star formation within a clumpy outflow and star formation induced by outflow shocks--- can explain the ubiquitous distribution of young stars in the ANGST footprint. Launched by the outflow and/or stripped from the M82 disk, HI and molecular gas exist throughout the ANGST footprint (see Figure \ref{fig:sfh-viz} and \citealt{Leroy2015})--- fuel for the stars that will come to populate the stellar halo.

The story of the origin of young halo stars put forth in this work involve highly non-linear, multi-phase gas physics that is inherently difficult to model. These are mechanisms governed by a range of physical parameters such as M82's starburst history, efficiency of supernova feedback, wind mass-loading, the galaxy’s dynamical mass distribution, and the nature of its environment, including the composition, density, and morphology of clouds of stripped gas. Many of these factors are challenging to constrain observationally, making it difficult to build realistic simulations that can favor or rule out the proposed story on theoretical grounds alone. 

Nevertheless, there exists a promising observational approach for confirming or ruling out the ram pressure stripping scenario. If the stars in the M82 Tail, indeed formed in this way, they must have systematically lower line-of-sight velocities compared to M82. This is because the gas they would have formed from would have experienced a deceleration from ram-pressure forces compared to the bulk of stars in the M82 disk. These velocity measurements could be achieved through spectroscopic follow-ups of bright stars along the M82 Tail. Spectroscopic follow-ups would also allow metallicity measurements. As \citet{Rao2025} suggest, these measurements in the Southern Arcs could help discriminate between degenerate star forming processes. In addition, SFH measurements of M82’s disk from deep NIRCam observations would help constrain the galaxy’s starburst history, while high-quality recent SFHs of halo stars near the disk could elucidate the processes that shape star formation in M82’s inner halo.

Beyond M82, widespread extraplanar star formation may be a general feature of systems hosting powerful starburst or AGN outflows. The challenges for observationally testing this hypothesis, however, are numerous. The first is proximity. There are a limited number of galaxies hosting outflows in the Local Universe ($D\lesssim10\,$Mpc), which are accessible to depth-limited resolved star studies of their halos. The farther we go, the harder star-galaxy separation becomes with large field-of-view (FOV) ground-based telescopes--- necessary for making panoramic maps of stellar halos and determining regions for deeper, focused, space-based follow-ups.

Even among nearby (D$\lesssim10\,$Mpc) galaxies, there are other limitations. The galaxy needs to be edge-on and must have an outflow that is sufficiently powerful for its size, if we are to be able to tell apart halo star formation from regular disk star formation. In the starburst galaxy NGC 253, for example, the outflow is not sufficiently powerful and does not visually make it beyond the optical disk of the galaxy. Consequently, HST fields in the halo do not pick up any significant young populations \citep{Monachesi2016, Harmsen2017}. There are, however, several nearby dwarf starburst galaxies with outflows such as NGC 1569, NGC 4449, and NGC 1705 \citep[e.g.,][]{Martin1999, Cignoni2018}. The challenge with these systems is that they are quite irregular, making outflow-induced star formation potentially challenging to separate from star formation in the overall galaxy.

The upcoming Nancy Grace Roman Space Telescope, with its $\sim0.3$ square degrees FOV, will vastly improve our capability of accurately mapping large swaths of $D\lesssim10\,$Mpc galaxy stellar halos \citep[e.g.,][]{Williams2020,Aganze2024}, potentially making it easier to detect young halo populations with improved star-galaxy separation. While Roman provides the breadth needed for nearby halos, JWST NIRCam offers the sensitivity and resolution to push into more distant regimes allowing us to target starburst systems like NGC 3079 ($D\sim16\,$Mpc) and NGC 3628 ($D\sim11\,$Mpc). The limitation here is that young stellar populations are fainter in the near-infrared filters used by Roman and JWST. These filters also give narrower color baselines than in the optical, which makes it harder to tell young populations apart from the older ones. Nevertheless, these facilities will allow us to approach hypothesis-testing on a statistical footing, enabling us to better understand how a galaxy can co-evolve with its surrounding environment.

\begin{acknowledgments}
We thank the anonymous referee for their insightful comments, which have improved our discussion. We also thank Paul Price for his work on the Subaru data reduction and acknowledge useful conversations with Sean D. Johnson. This work was partly supported by HST grant GO-16185 and JWST grant GO-5145 provided by NASA through a grant from the Space Telescope Science Institute, which is operated by the Association of Universities for Research in Astronomy, Inc., under NASA contract NAS5-26555. We acknowledge support from the National Science Foundation through grant NSF-AST 2007065. AS is supported by NASA through the Hubble Fellowship grant HST-HF2-51567 awarded by STScI. AM acknowledges support from the FONDECYT Regular grant 1212046, from the ANID BASAL project FB210003, and funding from the HORIZON-MSCA-2021-SE-01 Research and Innovation Programme under the Marie Sklodowska-Curie grant agreement number 101086388. 

This research is based on observations made with the NASA/ESA \textit{Hubble Space Telescope} General Observer program 16185. This work is partly based on observations utilizing the Pan-STARRS1 Survey. The Pan-STARRS1 Surveys (PS1) and the PS1 public science archive have been made possible through contributions by the Institute for Astronomy, the University of Hawaii, the PanSTARRS Project Office, the Max-Planck Society and its participating institutes, the Max Planck Institute for Astronomy, Heidelberg and the Max Planck Institute for Extraterrestrial Physics, Garching, The Johns Hopkins University, Durham University, the University of Edinburgh, the Queen’s University Belfast, the Harvard-Smithsonian Center for Astrophysics, the Las Cumbres Observatory Global Telescope Network Incorporated, the National Central University of Taiwan, the Space Telescope Science Institute, the National Aeronautics and Space Administration under grant No. NNX08AR22G issued through the Planetary Science Division of the NASA Science Mission Directorate, the National Science Foundation grant No. AST-1238877, the University of Maryland, Eötvös Loránd University (ELTE), the Los Alamos National Laboratory, and the Gordon and Betty Moore Foundation. This work is also based on observations obtained at the Subaru Observatory, which is operated by the National Astronomical Observatory of Japan, via the Gemini/Subaru Time Exchange Program. We thank the Subaru support staff--- particularly Akito Tajitsu, Tsuyoshi Terai, Dan Birchall, and Fumiaki Nakata--- for invaluable help preparing and carrying out the observing run. 

The authors acknowledge the significant cultural importance and enduring reverence of the summit of Maunakea to the Indigenous Hawaiian community since time immemorial. We are most fortunate to have the opportunity to conduct observations from this mountain.

\end{acknowledgments}

\begin{contribution}
VVR was responsible for the analysis, development of the scientific interpretations, and writing. EFB and AS were responsible for project ideation and mentoring, and were crucial for discussing scientific interpretations. AS conducted the HST observations, offered guidance on SFH fitting, and provided the Cibola dataset. EFB conducted the Subaru observations and performed the star-galaxy separation for the Subaru dataset. EB performed the cross-matching of the ANGST and Cibola datasets and contributed to writing. BW performed the HST photometry and ASTs. All other authors contributed towards project execution, review, and editing.


\end{contribution}

%

\facilities{HST, Subaru, VLA (NRAO), Mikulski Archive for Space Telescopes (MAST)}


\software{numpy \citep{numpy}, matplotlib \citep{matplotlib}, BEAST \citep{Gordon2016}, astropy \citep{astropy}, DOLPHOT \citep{Dolphin2000PHOT}, MATCH \citep{Dolphin2002, Dolphin2012, Dolphin2013,Dolphin2016}}
\appendix

\section{Completeness and Bias of the Photometry}\label{appendix:comp-bias}
\begin{figure*}
    \centering
    \adjustbox{max height=\dimexpr\textheight-3\baselineskip\relax}{%
        \includegraphics{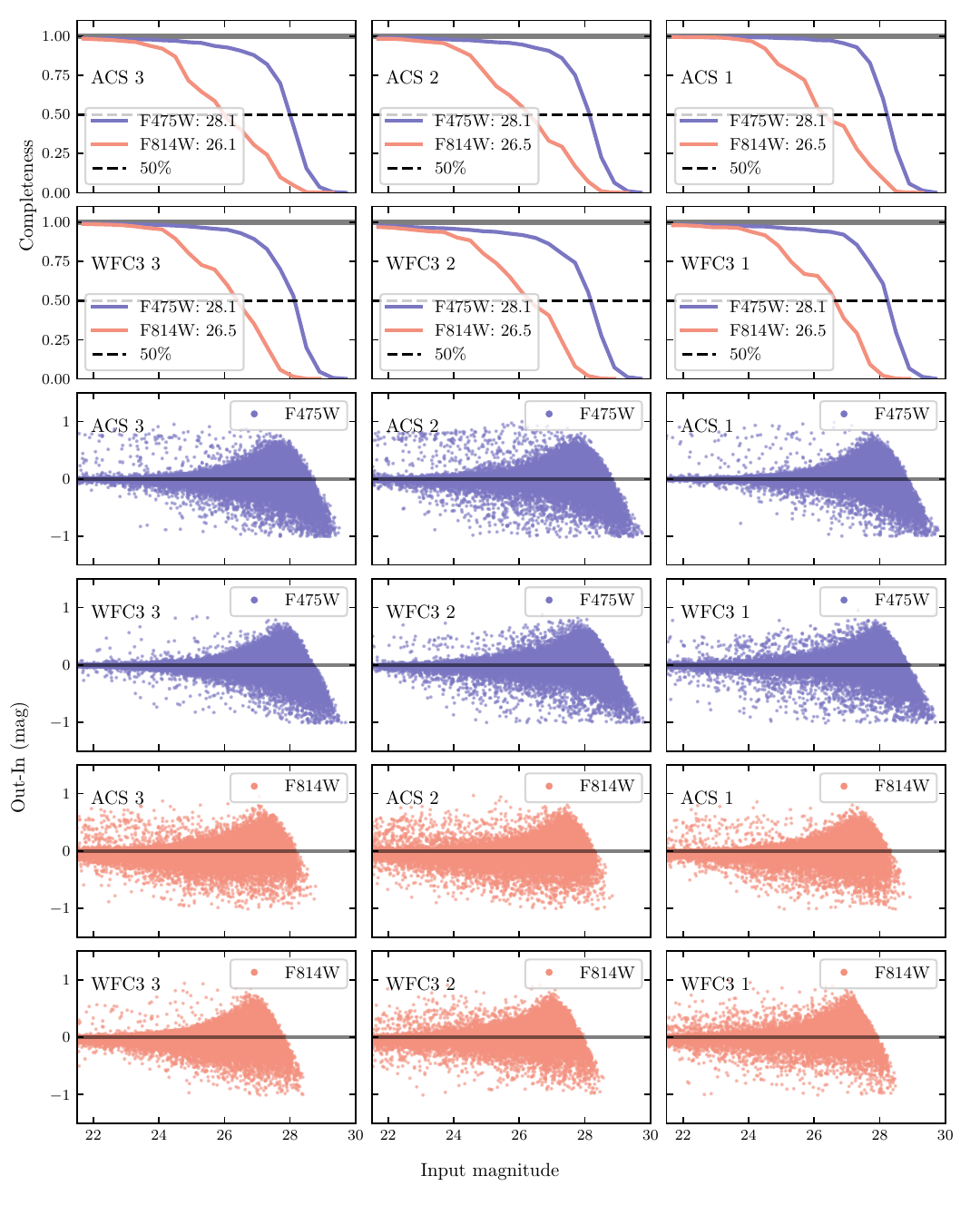}
    }
    \caption{Completeness of photometry in the Deep Halo Fields as a function of input magnitude (\textbf{Top}) and photometric bias or error in recovered magnitudes as a function of input magnitude for the F475W (\textbf{Middle}) and F814W (\textbf{Bottom}) filters from artificial star tests (ASTs). The subplots for each set have been arranged spatially.}
    \label{fig:hst-deep-completeness-bias}
\end{figure*}

In Figure \ref{fig:hst-deep-completeness-bias}, we show the completeness and bias of the photometry of all the Deep Halo Fields, resulting from artificial star tests. 

\section{Uncertainties in Star Formation Histories}
Uncertainties in our SFH analysis arise from both random and systematic uncertainties. Random uncertainties stem from the finite number of stars, and therefore generally scale inversely with this number. We calculated random uncertainties using a hybrid Monte Carlo(MC) process \citep{Duane1987} implemented through \texttt{MATCH} \citep[see][for implementation details]{Dolphin2013}. The result of this Markov Chain MC routine is a sample of 10,000 SFHs whose density in parameter space is proportional to its probability density. Upper and lower random error bars are calculated by identifying the boundaries of the region containing 68\% of the samples.

For systematic uncertainties, we recalculated the SFHs using three different sets of models: PADUA, BASTI, and MIST. Differences between these models are most pronounced in the post-main sequence phases of stellar evolution and for massive stars, primarily due to variations in modeling convection, mass loss, and rotation \citep{Conroy2013}. For example, the systematic difference in $\sim250-630\,$Myr SFHs in ACS Field 2 between PADUA and BASTI/MIST arises from the luminosities of HeB stars being marginally brighter in BASTI/MIST compared to PADUA.

\section{Validating the Star Formation Histories}
\begin{figure*}
    \centering
    \includegraphics[width= \linewidth]{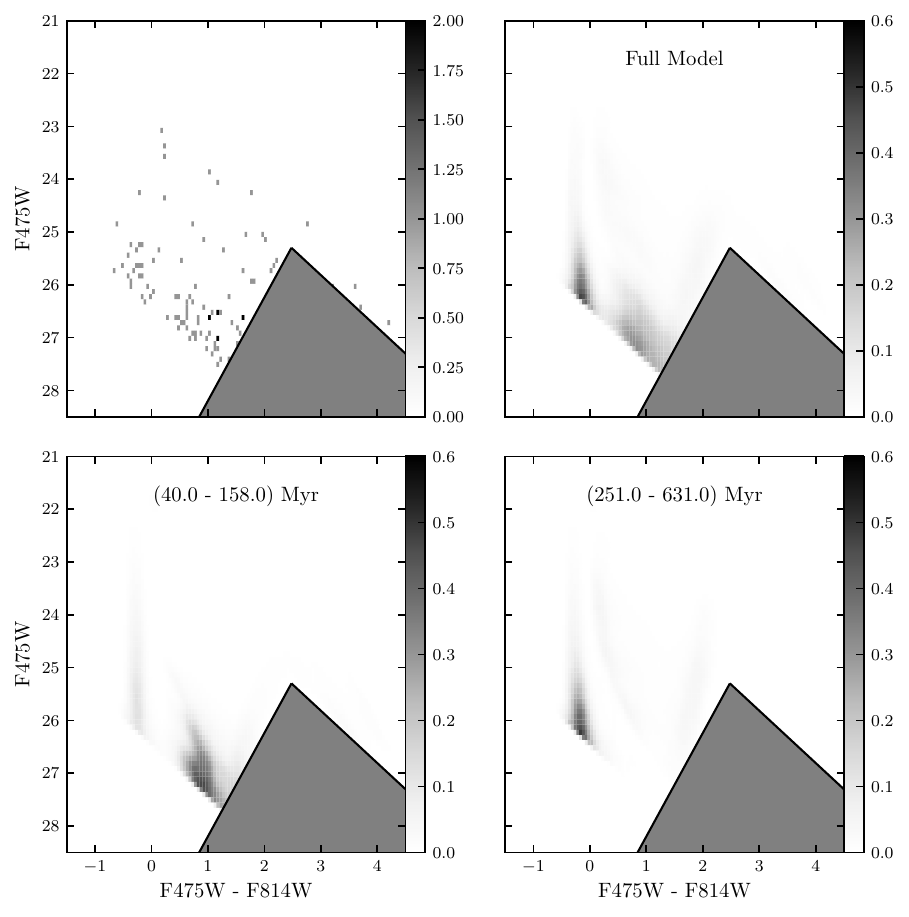}
    \caption{(\textbf{Top Left}) Observed Hess diagram of WFC3 Field 2. (\textbf{Top Right}) Best-fit PADUA model Hess diagram for WFC3 Field 2. (\textbf{Bottom Left}) Model Hess diagram generated with $40-158\,$Myr SFRs removed from the best-fit SFH. (\textbf{Bottom Right}) Model Hess diagram generated with $251-631\,$Myr SFRs removed from the best-fit SFH.}
    \label{fig:isetz-test}
\end{figure*}

Given the relatively small number of stars we use for obtaining SFH fits in the Deep Halo Fields, we attempt to validate our SFH results. Since much of our discussion centers on stars formed $\gtrsim 250\,$Myr and $\lesssim250\,$Myr, the goal of this exercise is to check whether the SFRs reported in these epochs actually correspond to physical features in the observed Hess diagram. Using WFC3 Field 2 as an example, we remove the $40-158\,$Myr and $251-631\,$Myr time bins from its SFH and generate a model Hess diagram with an $A_V$ of 0.05 and $dA_V$ of 0.4 using the PADUA models (Figure \ref{fig:isetz-test}). We can clearly see for each of these two cases that the model Hess diagrams with the modified SFHs miss important features present in the observed Hess diagram. The model Hess diagram with the $40-158\,$Myr SFRs removed fails to recover stars at the bottom of the MS and some of the bright HeBs whereas the model with the $251-631\,$Myr SFRs removed fails to recover most of the faint, red HeBs. This confirms that the SFH fitting procedure employed in this work generates SFRs based on actual physical features in the observed CMD.

\bibliography{main}{}
\bibliographystyle{aasjournalv7}



\end{document}